\documentclass[12pt,preprint]{emulateapj} 
\usepackage{lscape}

\begin{document}
\slugcomment{Accepted by AJ}

\title{Seventy-One New L and T Dwarfs from the Sloan
Digital Sky Survey}
\author{K. Chiu\altaffilmark{1},
X. Fan\altaffilmark{2},
S. K. Leggett\altaffilmark{3},
D. A. Golimowski\altaffilmark{1},\\
W. Zheng\altaffilmark{1},
T. R. Geballe\altaffilmark{4},
D. P. Schneider\altaffilmark{5},
J. Brinkmann\altaffilmark{6}}

\altaffiltext{1}{
Department of Physics and Astronomy, The Johns Hopkins University,
   3400 North Charles Street, Baltimore, MD 21218, USA }
\altaffiltext{2}{Steward Observatory, The University of Arizona, Tucson, AZ 85721}
\altaffiltext{3}{United Kingdom Infrared Telescope, Joint Astronomy Center,
660 North A'ohoku Place, Hilo, Hawaii 96720}
\altaffiltext{4}{Gemini Observatory, 670 North A'ohoku Place, Hilo, HI 96720}
\altaffiltext{5}{Department of Astronomy and AStrophysics, Pennsylvania State University, 
525 Davey Laboratory, University Park, PA 16802}
\altaffiltext{6}{Apache Point Observatory, P.O. Box 59, Sunspot, NM 88349}

\begin{abstract} We present near-infrared observations of 71 newly discovered L and T dwarfs,
selected from imaging data of the Sloan Digital Sky Survey (SDSS) using the $i$-dropout technique.  
Sixty-five of these dwarfs have been classified spectroscopically according to the near-infrared L dwarf 
classification scheme of Geballe et al. and the unified T dwarf classification scheme of Burgasser 
et al.  The spectral types of these dwarfs range from L3 to T7, and include the latest 
types yet found in the SDSS.  Six of the newly identified dwarfs are classified as early- to mid-L dwarfs according to 
their photometric near-infrared colors, and two others are classified photometrically as M dwarfs.  We also present 
new near-infrared spectra for five previously published SDSS L and T dwarfs, and one L dwarf and 
one T dwarf discovered by Burgasser et al.\ from the Two Micron All Sky Survey.  
The new SDSS sample includes 27 T dwarfs and 30 dwarfs with spectral types spanning the complex L--T transition (L7--T3).  We 
continue to see a large ($\sim$ 0.5 mag) spread in $J$--$H$ for L3 to T1 types,  and a similar spread 
in $H$--$K$ for all dwarfs later than L3.  This color dispersion is probably due to a range of grain 
sedimentation properties, metallicity, and gravity.  We also find L and T dwarfs with unusual colors 
and spectral properties that may eventually help to disentangle these effects.

\end{abstract}

\section{Introduction}

Over the last decade, the search for rare astronomical objects has undergone an explosion of
productivity, thanks in part to the specialized labor associated with modern digital sky surveys.
Innovative programmers and instrumentalists designing highly automated software pipelines and 
large imaging arrays have laid the groundwork for massive databases that can now be mined for 
objects once only thought to exist.   The result of these efforts has been the most productive 
period in history for the discovery of rare objects, which range from the brightest and most 
distant quasars \citep{fan99,fan00a,fan01a,fan01b,fan03,fan04,zheng00} to the 
faintest and nearest stars and brown dwarfs (\citealt[hereafter G02]{k99,burg99,k00,fan00b,
leggett00,burg02a,g02}; \citealt[hereafter K04]{k04}). 

\begin{figure*}[t]
\epsscale{0.8}
\figurenum{1}
\plotone{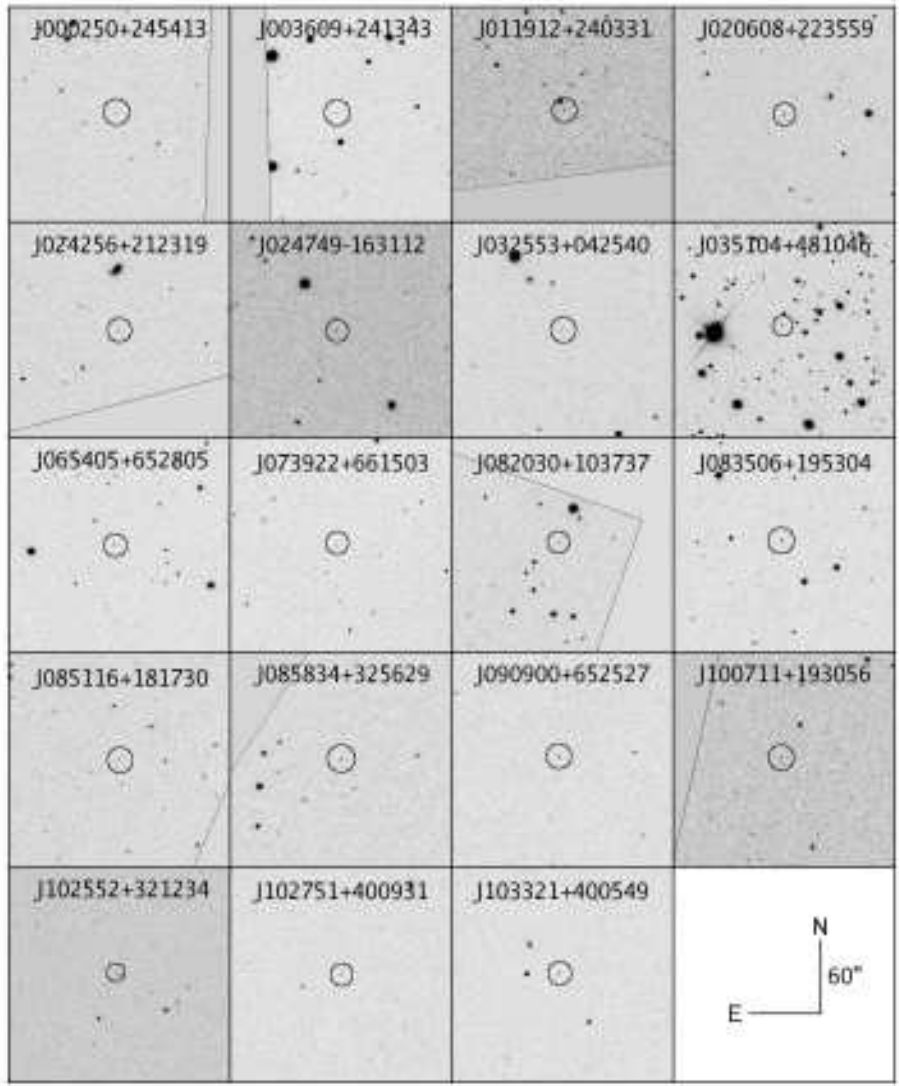}
\caption{SDSS $z$ finding charts for the 73 new M, L, and T dwarfs found in this work. North is up, East is left.  The fields of view are $3' \times 3'$.  (Full resolution versions of these finding charts will appear in the published journal edition.)}
\end{figure*}

\begin{figure*}[t]
\epsscale{0.8}
\figurenum{1}
\plotone{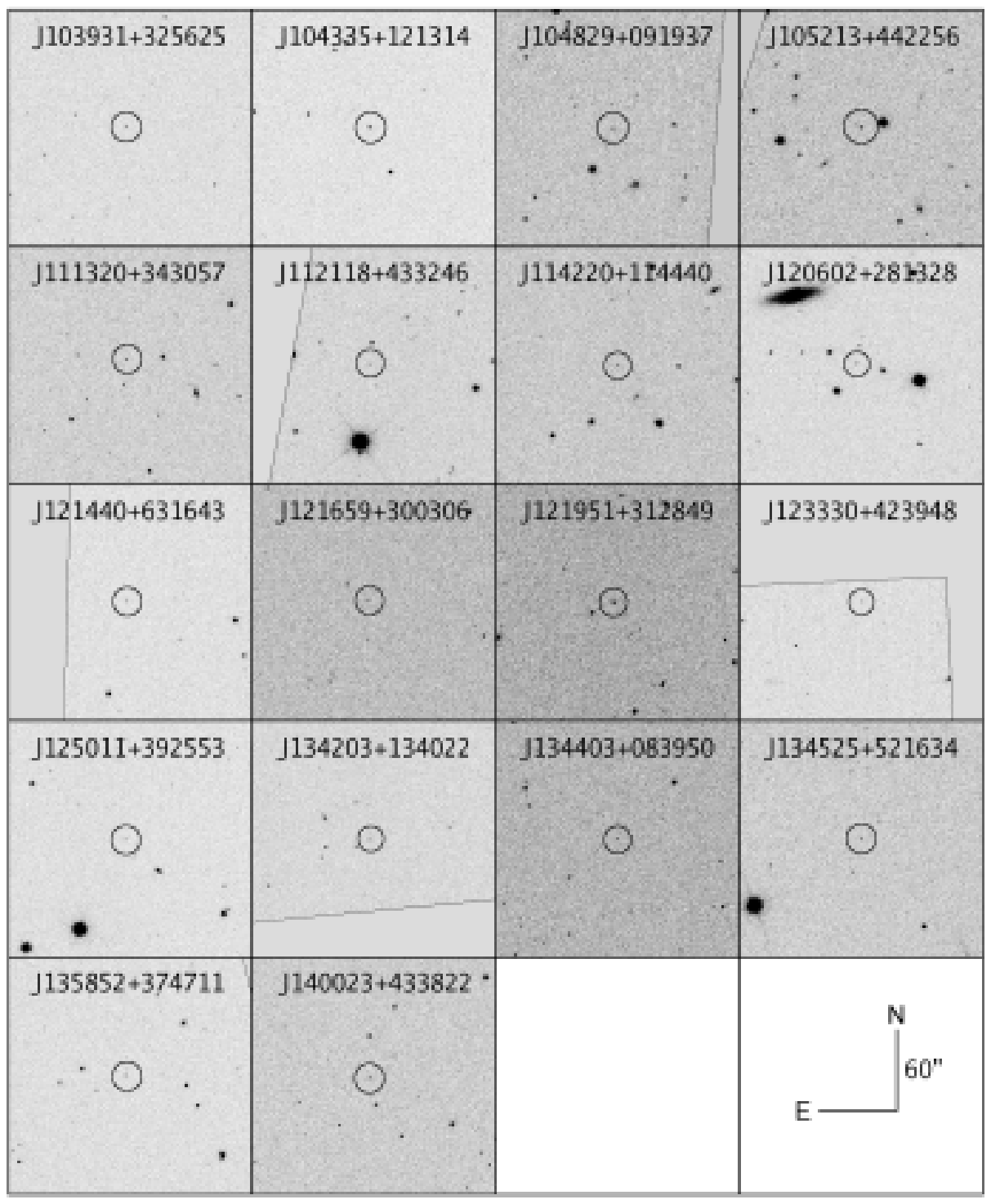}
\caption{(continued) SDSS $z$ finding charts for the 73 new M, L, and T dwarfs found in this work. North is up, East is left.  The fields of view are $3' \times 3'$.   (Full resolution versions of these finding charts will appear in the published journal edition.) }
\end{figure*}

\begin{figure*}[t]
\epsscale{0.8}
\figurenum{1}
\plotone{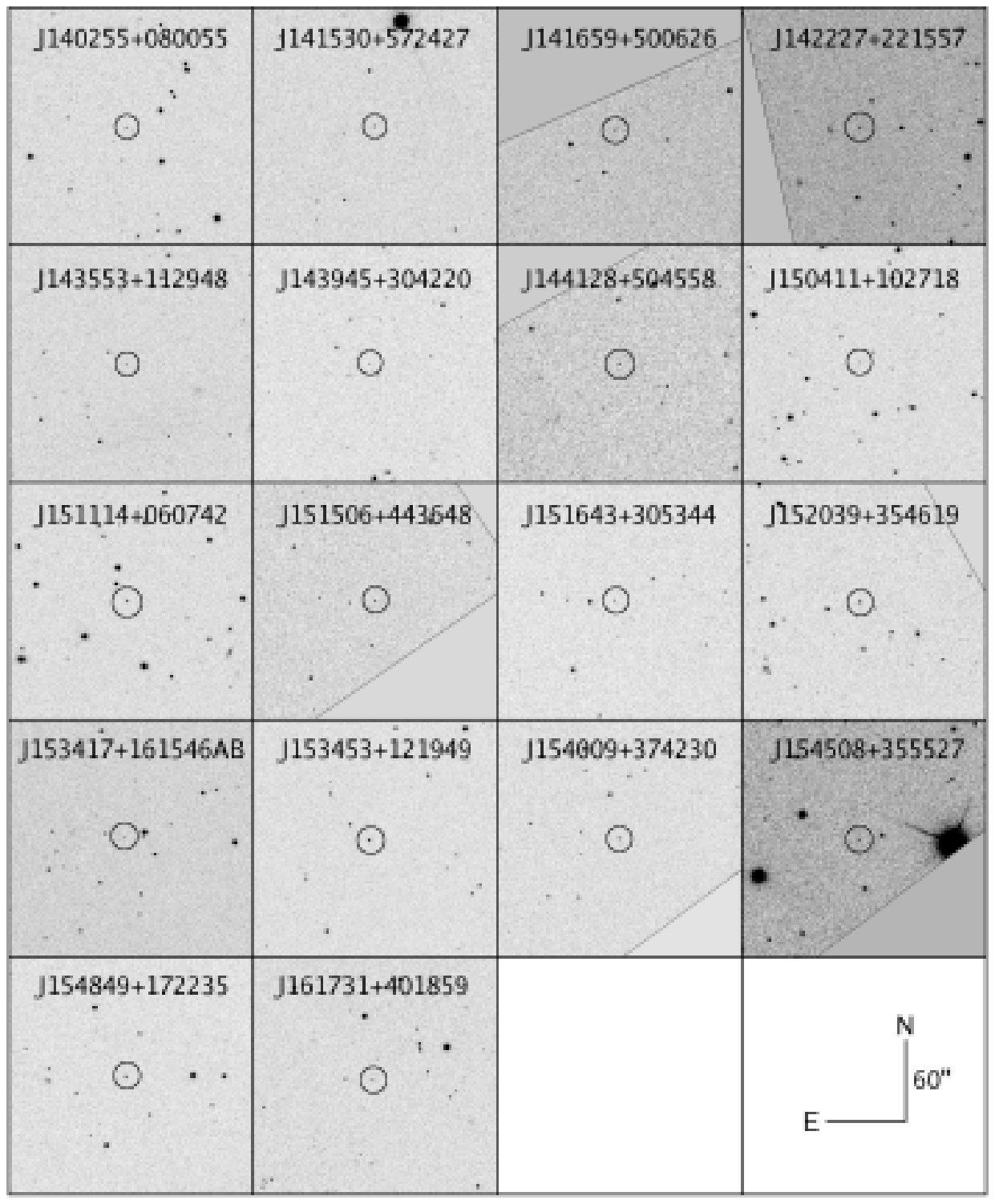}
\caption{(continued) SDSS $z$ finding charts for the 73 new M, L, and T dwarfs found in this work. North is up, East is left.  The fields of view are $3' \times 3'$. (Full resolution versions of these finding charts will appear in the published journal edition.)}
\end{figure*}

\begin{figure*}[t]
\epsscale{0.8}
\figurenum{1}
\plotone{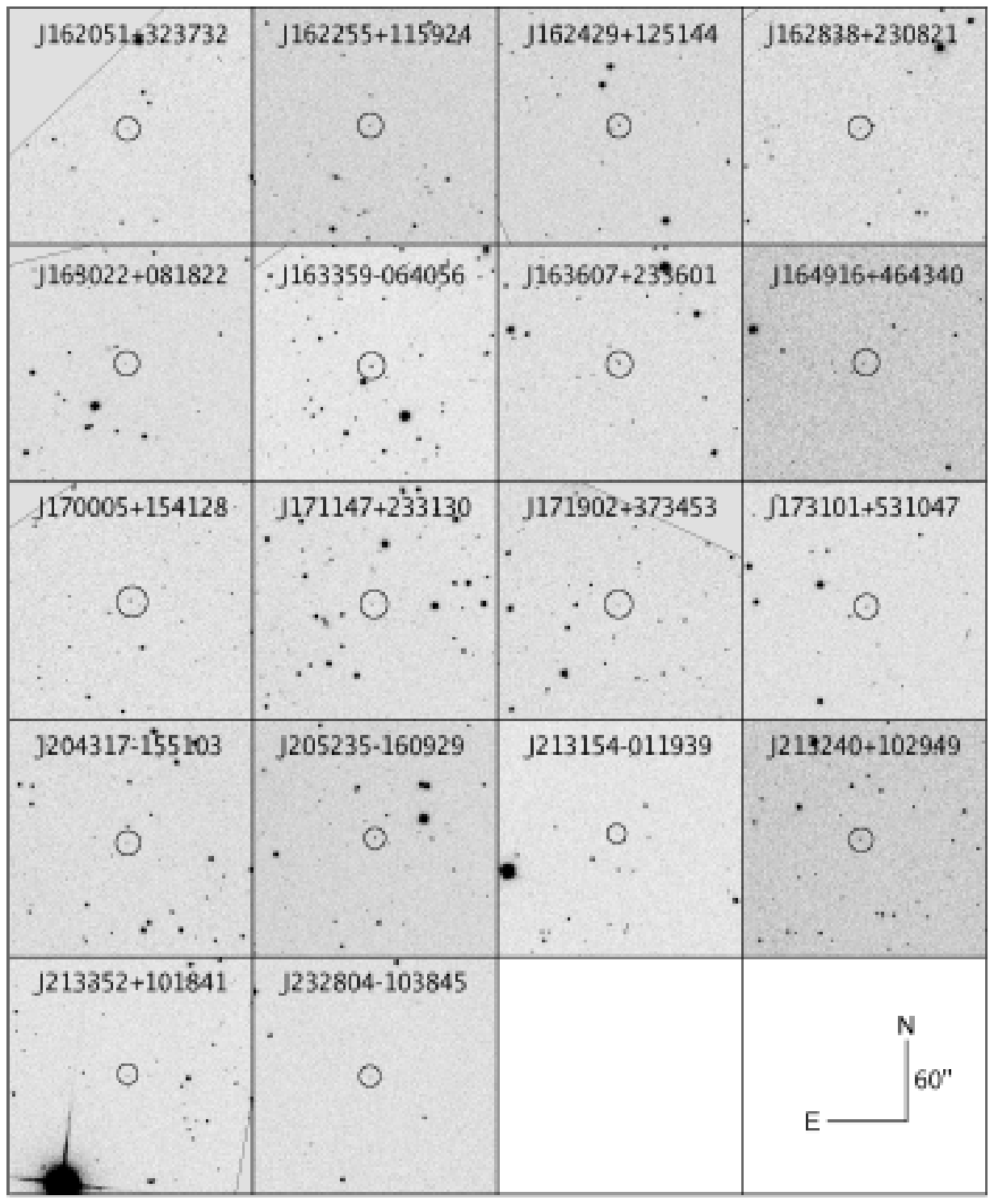}
\caption{(continued) SDSS $z$ finding charts for the 73 new M, L, and T dwarfs found in this work. North is up, East is left.  The fields of view are $3' \times 3'$.  (Full resolution versions of these finding charts will appear in the published journal edition.)}
\end{figure*}

\begin{figure}[h]
\epsscale{1.1}
\figurenum{2}
\plotone{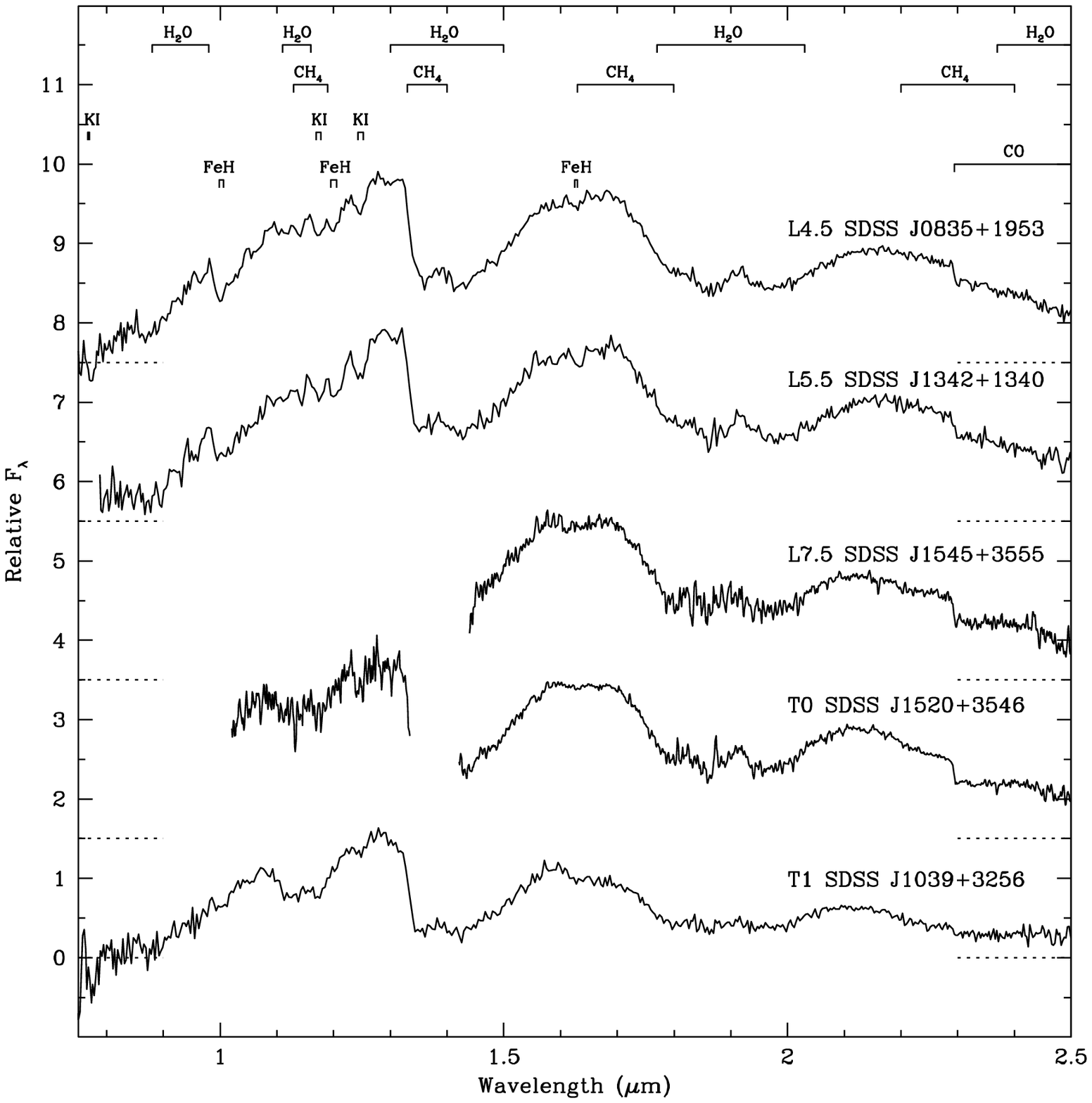}
\end{figure}

\begin{figure}[h]
\epsscale{1.1}
\figurenum{2}
\plotone{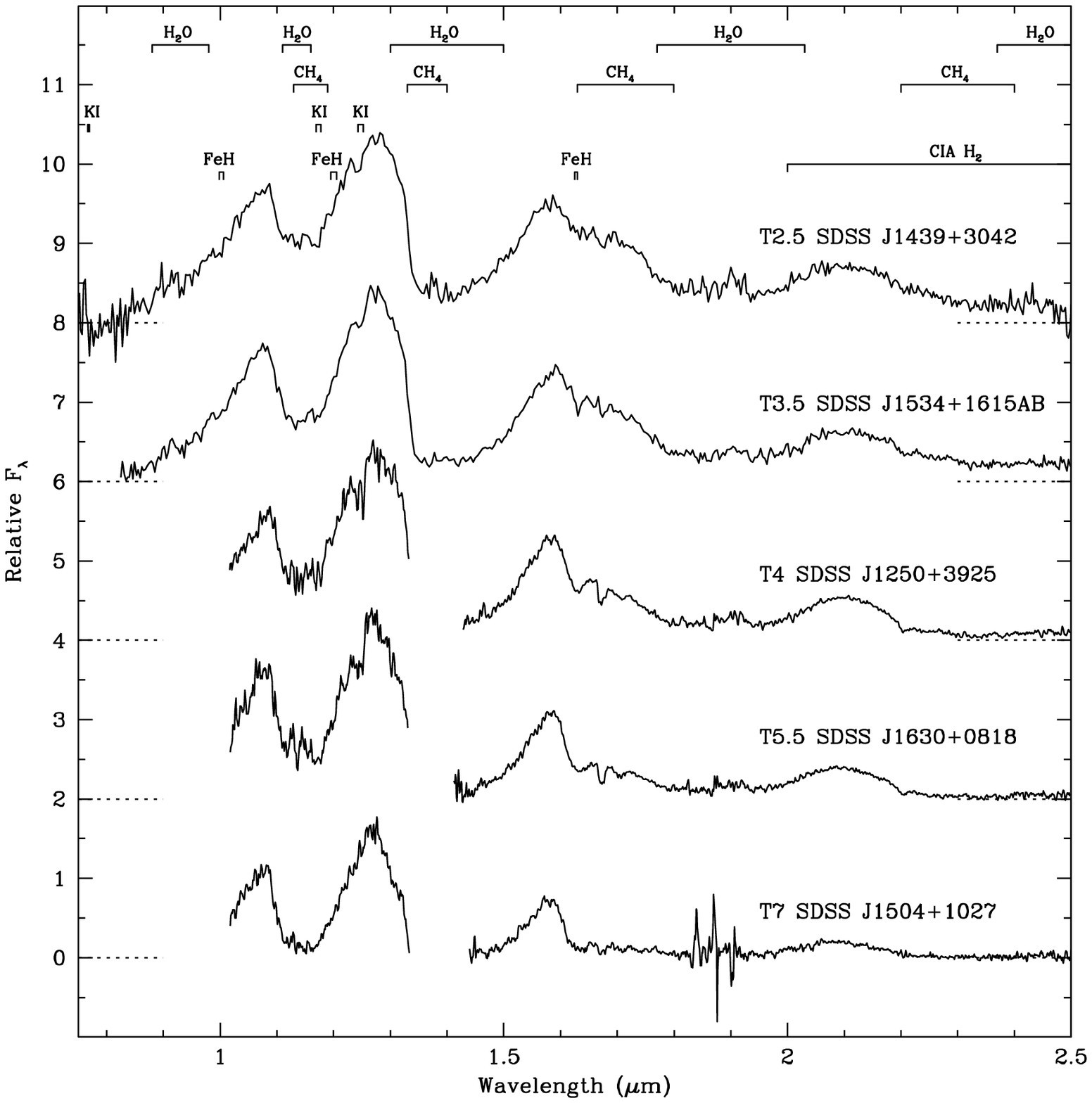}
\caption{Representative optical and near-infrared spectra of the L and T dwarfs discovered in this work.  The typical spectral resolution is $\sim150$.   SDSS designations are given above the dashed line corresponding to the zero flux level.  The locations of significant spectral features are indicated at the top of each panel.  
}
\end{figure}

Since the first discoveries of L and T dwarfs in the early 1990s, the ranks of these spectral 
classes have been populated through systematic searches of large-area, optical and near-infrared 
surveys such as the Deep Near Infrared Survey of the Southern Sky \citep[DENIS;][]{epchtein97},
the Two Micron All Sky Survey \citep[2MASS;][]{skrutskie97}, and the Sloan Digital Sky Survey 
\citep[SDSS;][]{york}.  Although time intensive, these searches are more productive and 
efficient than the tedious, pointed surveys that preceded them.  The optical and near-infrared
spectra of the many known L and T dwarfs have yielded well-defined classification schemes 
and valuable information about the chemical compositions, temperatures, and other physical 
characteristics of ultracool dwarfs (\citealt{k00,burg02a,burg05}; G02; K04, \citealt{gol04a}).

L and T dwarfs, which link the lowest mass stars and the highest mass planets, are the subjects 
of a broad range of observational and theoretical studies \citep{k05}.  A principal goal of these studies is 
the determination of the substellar mass function, which is fundamental to understanding the star
formation process at the lowest mass end and how it is related to the formation and evolution
of planetary systems \citep{burg04b,allen05}.  The hundreds of L dwarfs and more than 50 T dwarfs discovered from 2MASS 
and SDSS data compose a statistical sample of brown dwarfs essential to understanding the 
demographics and eventually the luminosity and mass functions of substellar objects (Fan et al., in prep.). 

Since 1999, we have used SDSS imaging data to select L and T dwarf candidates and carry out 
follow-up observations of their photometric and spectroscopic properties (\citealt{strauss99,
tsvetanov00,fan00b,leggett00,leggett02,schneider02}; G02; \citealt{hawley}; K04; \citealt{gol04a}).
The SDSS $i$--$z$ color, which is our primary selection criterion, is sensitive to brown dwarfs 
of all effective temperatures as long as they are bright enough to be detected in the $z$ band.
Near-infrared selection criteria, such as those employed by the 2MASS group \citep[e.g.,][]{burg02a},
are more sensitive to late-T dwarfs which are optically faint, but they are less efficient in 
detecting dwarfs at the L--T transition because their near-infrared colors are indistinguishable 
from those of the more numerous M dwarfs.  Therefore, the SDSS selection criterion is better suited
for defining a complete, magnitude-limited sample of field brown dwarfs.  

In this paper, we report the discovery and properties of 71 L and T dwarfs drawn from over 3500
deg$^2$ of SDSS imaging data.  We describe the techniques for identifying 
and confirming these dwarfs based on their SDSS and near-infrared colors.  We classify 65 of 
the L and T dwarfs from their near-infrared spectra, and we identify 6 L dwarfs and 2 M dwarfs
from their photometric colors.  The observations are presented in \S 2, spectral 
classification of the sample is presented in \S 3, and variations in the photometric and
spectroscopic  properties are discussed in \S 4.  Our conclusions are given in \S 5.

\section{Observations}

The search for nearby, ultracool dwarfs at optical wavelengths is fortuitously related to the 
search for the most distant galaxies and quasars in the universe.  Because both types of objects
have very red optical colors caused by rising flux into the near-infrared, a search for one yields the other (or from another perspective, 
one is the unwanted contaminant of the other).  The ``$i$-dropout'' and related techniques have 
been employed in nearly all high-redshift observing programs,  so a joint effort to find L and T 
dwarfs with these techniques benefits both programs.  

\subsection{General Selection Method}

We select candidate L and T dwarfs using the $i$ and $z$ magnitudes of sources listed in the 
SDSS photometric catalog.  The SDSS uses a dedicated 2.5~m telescope in a drift-scanning mode to acquire 
digital images.  The hardware and software pipelines that produce the final astrometry and 
photometry have been described by the project collaborators elsewhere in detail: \citet{fukugita}, \citet{gunn98}, \citet{hogg},
\citet{lupton}, \citet{stoughton}, \citet{smith}, \citet{pier}, and \citet{abz}.  

Fan et al. (2001b) 
described the $i$-dropout selection and identification procedures used in this program.  The photometric
selection criteria are:

\begin{equation} 
\begin{array}{l} 
z < 20.4, \\ 
\sigma(z) < 0.12, \\ 
i - z > 2.2.\\ 
\end{array} 
\end{equation}

Because most of the SDSS area is imaged only once and our $i$-dropout technique favors one-band detections, 
our initial list of L and T dwarf candidates is heavily contaminated by cosmic rays.  False detections 
from cosmic rays and intrinsic faintness are the main problems preventing the SDSS itself from discovering these objects, 
as well as $z>5.7$ quasars, in its automated spectroscopic oberving program \citep{richards02}.  
We conservatively remove cosmic rays from our list by visual inspection.   

The reality of the remaining bright candidates can be immediately
confirmed by correlating the list with the 2MASS All-Sky Point Source
  Catalog, which contains near-infrared photometry down to a $10\sigma$
  detection limit of $J=15.8$.  We matched our SDSS candidates with 2MASS
  sources, allowing for small astrometric errors caused by the proper
  motions of these nearby dwarfs.  Although the $z$--$J$ colors of ultracool
  dwarfs are very red ($z$--$J > 2$; Fan et al. 2001b), our fainter candidates
  often do not appear in the 2MASS catalog because of 2MASS's shallower
  imaging depth.  Those fainter candidates not confirmed by 2MASS were
  further examined through independent $J$-band imaging.  Additional
  SDSS $i$ and $z$ imaging was carried out to secure our primary color
  criterion of $i$--$z > 2.2$.

The followup $J$-band photometry allows us to distinguish very red L and T dwarf candidates from bluer high-redshift quasar 
candidates.  Quasars at redshifts $> 5.7$ have $z$--$J < 1.5$ because their intrinsic continua are 
dominated by a blue power law.  Ultracool dwarfs are more easily discovered than quasars because they are
much redder, and therefore, significantly brighter in the $J$ band.  Applying the selection criteria above to a 
6600 deg$^2$ survey area, 
we have confirmed 53 T dwarfs and over 100 L dwarfs (including those found in this work), but only 19 quasars with redshifts $> 5.7$.

Finally, we obtained near-infrared spectra of candidates whose colors matched
those of L and T dwarfs, at the
wavelength resolution needed to determine the dwarf classification indices.  

\subsection{Photometric Observations}

The area of sky covered in this work is 3526 deg$^2$, comprising
imaging runs recorded from early 2003 to 2005 and a few earlier runs that had been reprocessed through
the SDSS photometric pipeline.   Eighty four $i$-dropout candidates were selected 
using the criteria in Equation~(1), and then examined with the 2MASS catalog or follow-up $z$ and $J$ 
imaging.  A few additional candidates with attributes outside of our 
formal selection limits were also selected, in order to fill gaps during observations. 
 We obtained $z$-band images with the SPICAM imager on the Astrophysical Reseach Consortium 
(ARC) 3.5~m telescope at the Apache Point Observatory. The typical exposure time needed to verify the presence of a 
$z=20.4$ source was $\sim 60$~s.  We obtained $J$-band images with the $256 \times 256$ NICMOS 
camera on Steward Observatory's 2.3~m Bok Telescope, and the GRIM II and NIC-FPS imagers on the ARC 3.5~m telescope.  

Candidates surviving our initial photometric screening were then imaged through the Mauna Kea Observatory
(MKO) $J$, $H$, and $K$ bands \citep{sim02,tok02} using the United Kingdom Infrared Telecope (UKIRT) 
Fast-Track Imager \citep[UFTI;][]{ufti} or the NASA Infrared Telescope Facility (IRTF) imager/spectrometer 
SpeX \citep{spex}.  UKIRT Faint Standards \citep{haw01} were used to calibrate the data.  Typical exposure 
times were 60~s, with a five or nine point dither pattern.   

Table 1 lists the SDSS designations, SDSS $iz$ and MKO $JHK$ magnitudes, and observation details for the 73 
M, L, and T dwarfs discovered in this work.\footnote{The SDSS $iz$ magnitudes are based on the AB system, whereas the MKO 
$JHK$ magnitudes are based on the Vega system.  The photometric errors of these objects in the SDSS public data releases may differ from those published here due to reprocessing of the data.}   
Although the SDSS $i$-band $5\sigma$ detection limit is 22.5, the SDSS
  photometric pipeline yields asinh magnitudes below this flux limit
  (Lupton et al. 1999).  For this reason, we identify with brackets
  those SDSS $i$ magnitudes in Table 1 that imply non-detections.  
   The SDSS designations contain the J2000.0 right ascensions
and declinations in sexigesimal format, truncated to the significant digit shown.  The positions are accurate 
to better than 0\farcs1 in each coordinate.
For brevity, we hereafter refer to the dwarfs by the abbreviated form, SDSS JHHMM$\pm$DDMM.  One T dwarf, SDSS J1534+1615AB,  has recently 
been resolved into a close binary using the laser guide star and adaptive optics systems at the Keck Observatory \citep{liu06}.   
Figure 1 
displays $3'\times3'$ $z$-band finding charts for the 73 dwarfs in our sample.  

Table~2 lists MKO $JHK$ magnitudes for three 2MASS dwarfs (2MASS J0034+0523, 2MASS J1209$-$1004, 2MASS J2101+1756)
as well as UFTI $Z$ magnitudes for 2MASS J1209$-$1004 and seven previously published SDSS L and T dwarfs 
(\citealt{burg04a}; K04).  The $Z$-band data are not used in this analysis but are presented for 
future reference.

\subsection{Spectroscopic Observations}

We obtained near-infrared spectra of 65 newly identified SDSS dwarfs
  whose near-infrared colors are consistent with types mid-L and later.  We
  also obtained spectra of the T dwarf 2MASS J1209$-$1004 \citep{burg04a}, the
  mid-L dwarf 2MASS J2101+1756 (Kirkpatrick et al. 2000), three L dwarfs
  identified by K04 on the basis of their colors (SDSS J0740+2009,
  SDSS J0756+2314, and SDSS J0809+4434), one L dwarf with uncertain spectral
  type (SDSS J0805+4812; K04), and one T dwarf with incomplete spectral
  coverage (SDSS J2124+0100; K04).  The new spectra allow more accurate
  classification of these previously reported L and T dwarfs.  We did not
  obtain spectra of the two M dwarfs and six early- to mid-L dwarfs because
  of observing time constraints.

The spectra were obtained using the UKIRT Imager/Spectrometer \citep[UIST;][]{uist} and Cooled Grating 
Spectrometer \citep[CGS4;][]{cgs4} on UKIRT, as well as SpeX at the IRTF.  The UKIRT instrument configurations and 
observational procedures are described by K04.  SpeX was used in the low-resolution prism
mode with a slit width of 0\farcs5, giving a wavelength coverage of 0.8--2.5~$\mu$m with a resolution 
of $\sim 150$.  The on-target exposure time was 120~s, and each star was nodded along the slit for a 
total observation time of 48--80 min.  Flat fields and arc spectra for wavelength calibration were 
obtained using the SpeX calibration box lamps.  A0 stars were observed as telluric calibrators, and 
the data were reduced using the Spextool package \citep{spextool}.   The flux calibration of the 
spectra was improved by adjusting the relative flux level to match our $JHK$ photometry.  Tables~3 and 4
list the instruments and dates of observation for each dwarf in our spectral sample.

Figure 2 shows spectra of selected L and T dwarfs representing the range of types in our sample. 
\footnote{More extensive spectral and photometric data for this and other samples may be found at http://www.jach.hawaii.edu/$\sim$skl/LTdata.html}
Most spectra cover the wavelength range $\sim 1$--$2.5~\mu$m.  Interesting and diagnostic absorption 
features are identified above the displayed spectra. 

\section{Spectral Classification}

The near-infrared classification of L and T dwarfs is generally accomplished either by directly 
comparing the spectra with standard-dwarf templates or by measuring indices defined for their diagnostic 
spectral bands (i.e., H$_2$O and CH$_4$).  As in our past work, we use the spectral-index approach, 
which is robust even with medium to low signal-to-noise ratios and spectral resolution.  

To place our present sample of SDSS L dwarfs in the context of those presented by G02 and K04, we 
classified the L dwarfs using the near-infrared spectral indices of G02.  However, we classified the 
T dwarfs using the new spectral indices of \citet{burg05}, which unify the similar T classification 
schemes of G02 and \citet{burg02a}.  The unified scheme uses five primary indices: H$_2$O-$J$ centered 
at 1.15~$\mu$m, CH$_4$-$J$ at 1.32~$\mu$m, H$_2$O-$H$ at 1.4~$\mu$m, CH$_4$-$H$ at 1.65~$\mu$m, and CH$_4$-$K$ 
at 2.2~$\mu$m.   The values of the unified indices are inverted with respect to those of G02, i.e., the values
of the unified indices decrease with later T type.  We used both the G02 and unified indices to classify the 
suspected L--T transition dwarfs, as in some cases one scheme suggests a late-L type and the other scheme 
suggests an early-T type.   

Table 3 lists the G02 indices of 61 L and early-T dwarfs for which we have obtained new near-infrared spectra.
Of these, 56 are newly identified from the SDSS and 5 are previously known SDSS and 2MASS dwarfs, as described
in \S2.3.  Values in square brackets are uncertain and have not been included in the final spectral classification.  
When only a single index could be measured reliably, the spectra were visually compared with standard subtypes
for more accurate classification.  Table 4 lists the indices for 45 late-L and T dwarfs from the unified scheme of 
\citet{burg05}.\footnote{Revised spectral types for previously published SDSS T dwarfs and two late-L dwarfs 
(SDSS J1104+5548 and SDSS J2047$-$0718) are given in Table 14 of \citet{burg05}.}  In both tables, the uncertainty 
in the mean spectral type is $\pm 0.5$ subtype, unless otherwise stated.

Table 5 lists, in order of spectral type, the optical and/or near-infrared colors of the 72 dwarfs for which we have 
new spectra, plus 2MASS~J0034+0523.  Multiple photometric and spectroscopic measurements (including data from K04) were averaged
before determining the final colors and types.  The $i$--$z$ colors for dwarfs not detected in $i$ were computed using
the SDSS $5\sigma$ detection limit of $i=22.5$.  The spectral types of all newly discovered L and T dwarfs are taken 
from Tables~3 and 4, respectively.  The subtypes adopted for the L8--T1 dwarfs are those produced by the classification 
scheme that exhibited the more definitive and complete set of indices for its resultant subtype.  If both schemes 
indicated a late-L type, then the G02 designation was adopted.  If both schemes indicated an early-T type, then the 
unified T classification of \citet{burg05} was used.  The only exception to these rules is SDSS~J1511+0607 (T0~$\pm$~2),
for which a definitive set of indices was derived only in the G02 scheme.  The uncertainties of all the subtypes listed 
in Table~5 are $\pm 0.5$ subtype, unless otherwise stated.  

\begin{figure}[t]
\epsscale{1.0}
\figurenum{3}
\plotone{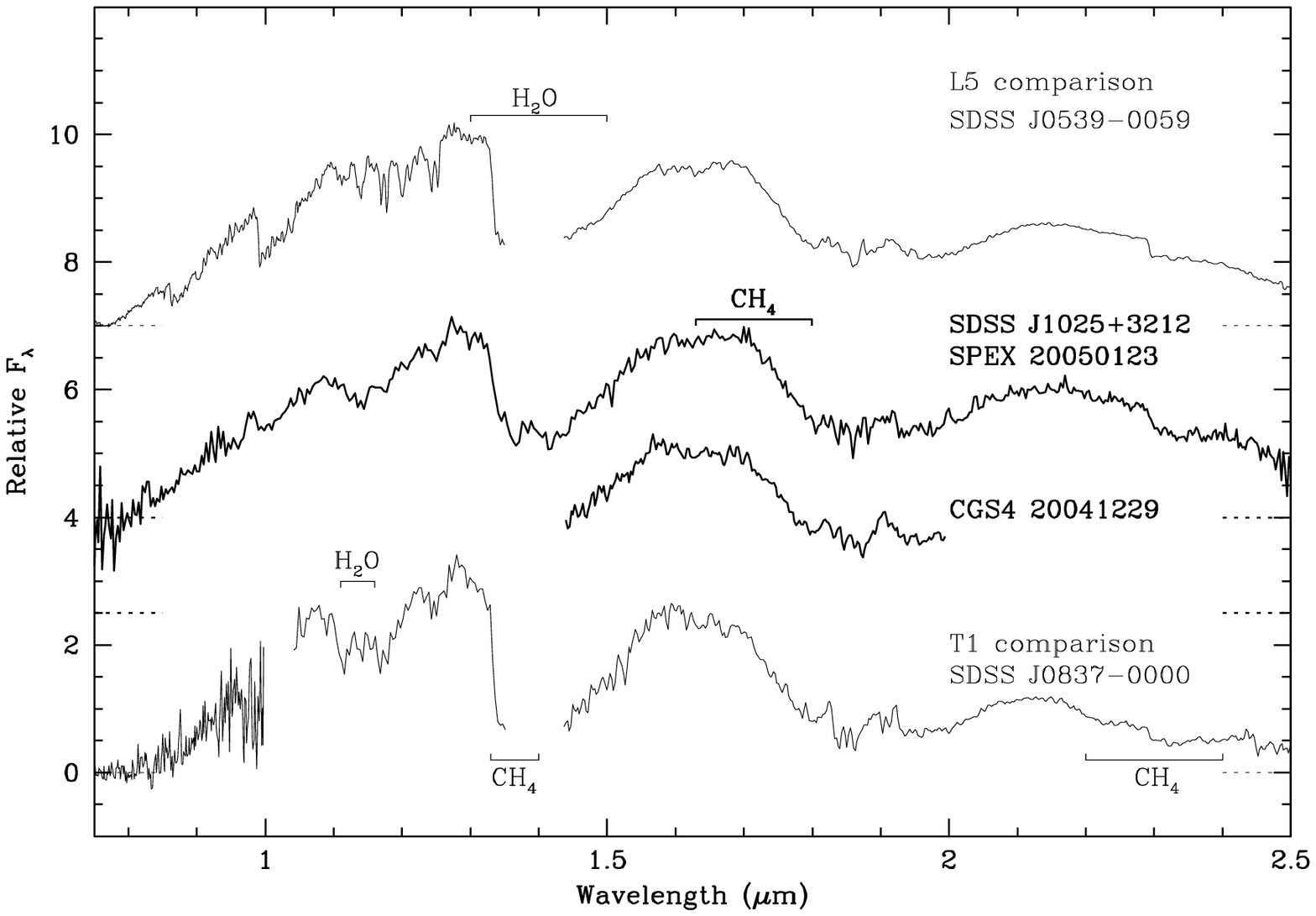}
\vspace*{-0.7in}
\caption{The unusual spectra of the L7.5~$\pm$~2.5 dwarf, SDSS J1025+3212 (thick lines).  The spectra show 1.1 $\mu$m H$_2$O and  1.4, 2.2 
$\mu$m CH$_4$ absorption features usually associated with T0 dwarfs, as well as H$_2$O features typical of L5 dwarfs.  The 1.6 $\mu$m CH$_4$ 
absorption features seen in the CGS4 and SpeX spectra appear variable.  Spectra of SDSS J0539$-$0059 (L5) and SDSS J0837$-$0000 (T1) are 
shown for reference \citep{leggett00}.  
}
\end{figure}

\begin{figure}[t]
\begin{center}
\epsscale{1.25}
\figurenum{4}
\hspace*{-0.3in}
\plotone{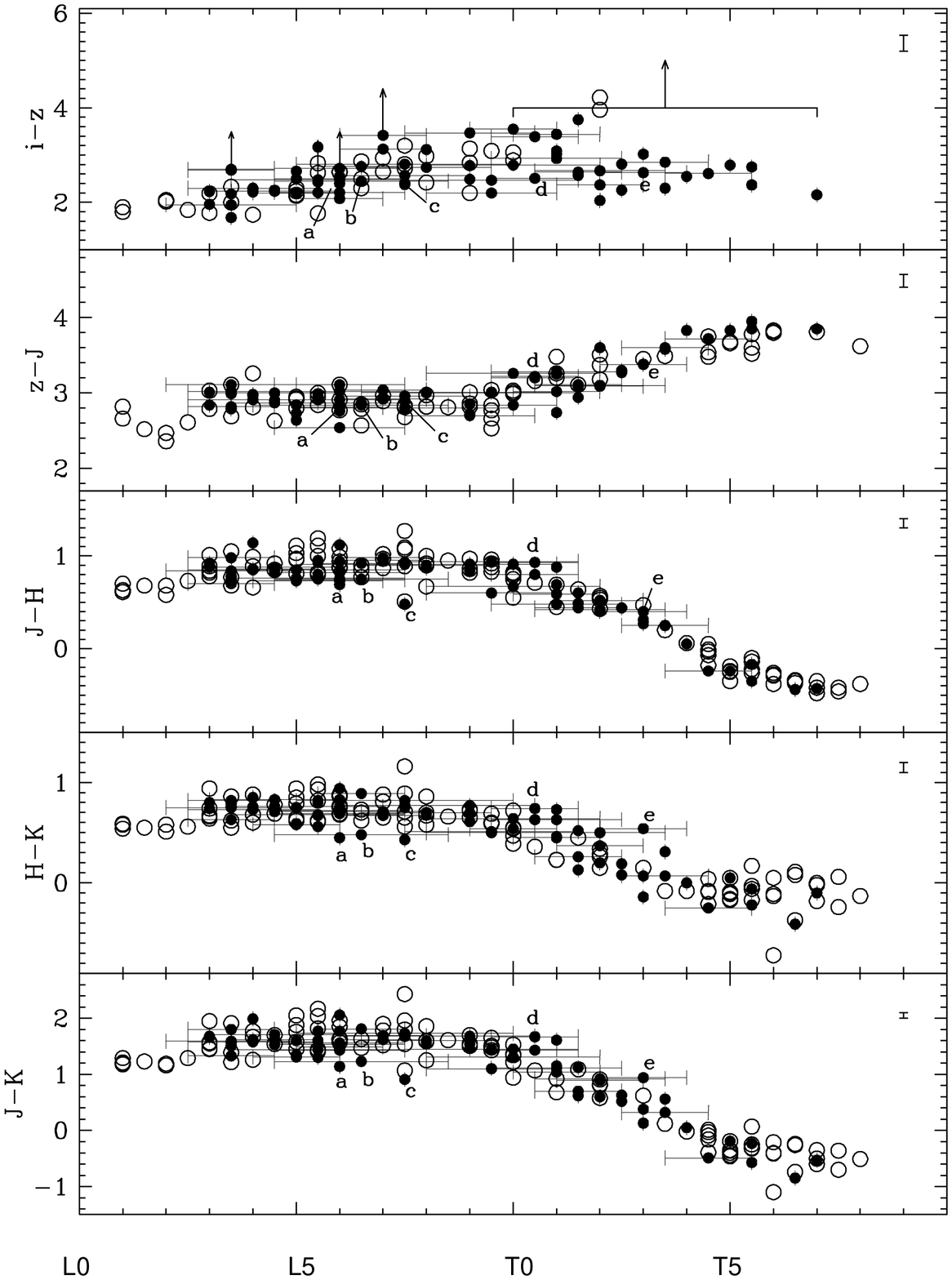}
\caption{Optical and near-infrared colors of the dwarfs listed in Table 5 of this work (solid points) and reported previously by \citet{k04} 
and \citet{leggett02} (open circles). Five dwarfs with unusual colors are labelled, as discussed in \S4: (a) SDSS J1033+4005, (b) SDSS J1422+2215, 
(c) SDSS J1121+4332, (d) SDSS J1516+3053, and (e) SDSS J1415+5724.  Horizontal error bars are shown for those dwarfs whose spectral types 
are more uncertain than $\pm 0.5$ subtype.  The small vertical error bar in the upper-rightmost corner of each panel indicates the typical 
color error.   In the top panel, the $i$--$z$ colors of dwarfs not detected in SDSS $i$ are shown with arrows as lower limits based on a $5\sigma$
detection limit of $i = 22.5$.  Dwarfs with type $>$T0 are almost all non-detections in SDSS $i$, shown by the large barred arrow in $i-z$.}
\end{center}
\end{figure}

The uncertainties of the L6--T1 dwarfs are typically $>1$ subtype, which suggests that the presence of condensate 
cloud decks undermines the internal consistency of the spectral indices.  Consequently, the indices 
become more a probe of cloud optical depth than a measure of effective temperature (\citealt{stephens},
K04, \citealt{leggett05}).  
Approximately 75\% of the dwarfs whose mean spectral types are very uncertain have CH$_4$-$K$
spectral types that are significantly earlier than indicated by their $J$- and $H$-band indices.
This effect is also seen in the L8--T0 sample of K04.  Because the optical and $K$-band 
fluxes emerge from more opaque regions of the atmosphere (specifically from regions above the 
cloud decks), they are likely to be better indicators of $T_{\rm eff}$ than the flux emerging 
from the clear $J$- and $H$-band windows.  \citet{gol04a} showed that $T_{\rm eff}$ is 
approximately constant for L7 to T4 types, so it might be expected that the L--T transition 
dwarfs have a constant optical and $K$-band spectral type of $\sim$~L8, while their $J$- and 
$H$-band indices reflect changes in optical depth caused by varying condensate clouds.

The newly identified L dwarf SDSS J1025+3212 (L7.5~$\pm$~2.5) has particularly scattered spectral 
indices (Tables~3 and 4), indicating that it may be multiple and/or variable.  Figure~3 shows the 
SpeX and CGS4 spectra of SDSS J1025+3212 along with comparison L5 and T1 spectra.  Its H$_2$O-$J$ 
index suggests a T0 type, while its CH$_4$-$J$, H$_2$O-$H$, and CH$_4$-$K$ indices suggest a 
mid-L type.  Evidence of CH$_4$ absorption at 1.6~$\mu$m is seen in the CGS4 spectrum but not in 
the SpeX spectrum obtained one month later.  Attempts to reproduce the spectrum as a composite of
L and T spectra failed.  High-resolution imaging and photometric monitoring are needed to investigate
the nature of this peculiar brown dwarf.

\begin{figure}[t]
\epsscale{1.1}
\figurenum{5}
\plotone{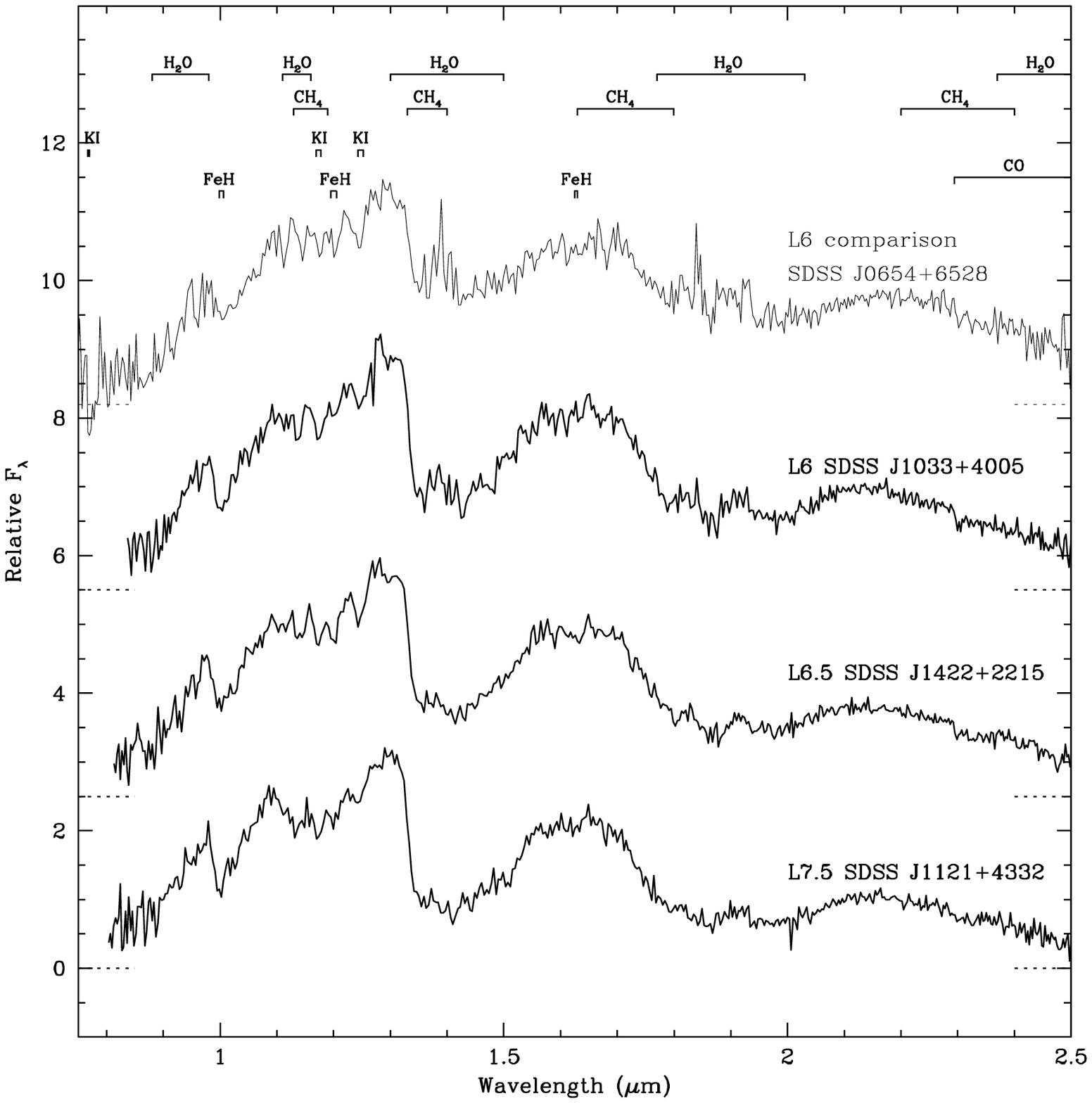}
\caption{Spectra of unusually blue L dwarfs with strong H$_2$O and  FeH absorption bands.  These features may be due to subsolar metallicity 
and/or thinner condensate cloud decks.  A spectrum of the typical L6 dwarf SDSS J0654+6528 in shown for reference.
 }
\end{figure}

\begin{figure}[t]
\epsscale{1.1}
\figurenum{6}
\plotone{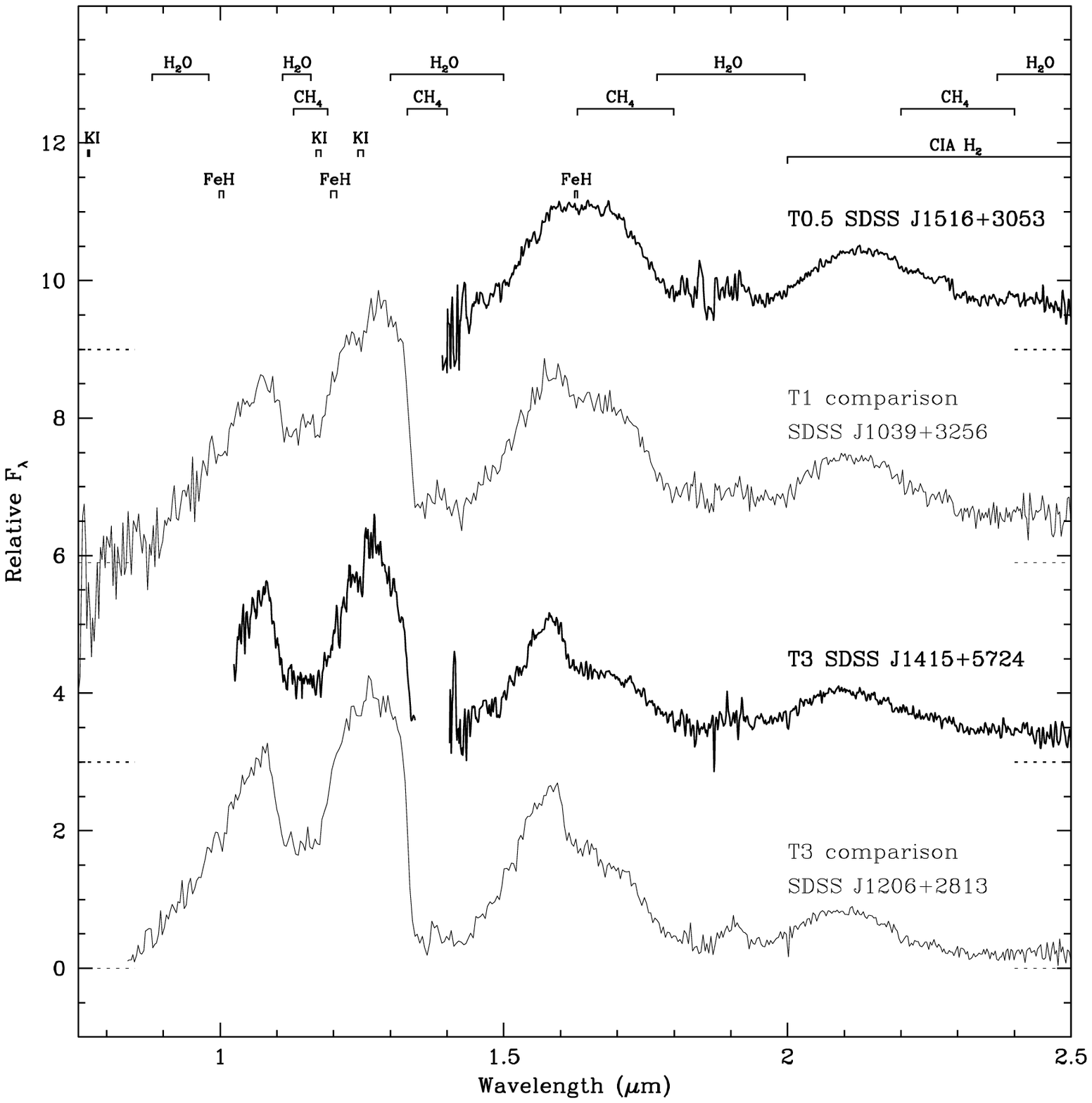}
\caption{Spectra of unusually red early T dwarfs (thick curves) compared with spectra of typical T dwarfs.  SDSS J1516+3053
has weak methane bands, but strong water bands.  SDSS J1415+5724 has unusually weak $K$-band methane.}
\end{figure}

Ambiguous types like SDSS J1025+3212 give cause for reexamining the near-infrared classification
scheme for L dwarfs, especially near the L--T transition.  Moreover, two of the first known early-T 
dwarfs, SDSS J0423-0414 and SDSS J1021-0304, have recently been found to be binary (Burgasser et al. 2006; 
Burgasser et al., in prep.). 
 These binary dwarfs contributed significantly to the definition of G02's standard 
indices at the L--T transition.  Revision of these indices must be based on single dwarfs,
so knowing which transition dwarfs are multiple is prerequisite to merging a revised L infrared classification
scheme with the unified T classification scheme of \citet{burg05}.  Several high-resolution imaging 
searches for close-binary brown dwarfs are being conducted with the {\it Hubble Space Telescope}
and ground-based adaptive optics imagers \citep{burg03,bou03,mcc04,gol04b,liu05,liu06,burg06}.  Segregating the 
single and multiple transition dwarfs will
not only help define the ``standard'' L and T sequences, but it will clarify our understanding of 
the breakup of the condensate cloud decks across the L--T transition, which is presently a 
controversial aspect of ultracool atmosphere models (\citealt{burg02b,tsuji03}; K04, Burrows et al. 2005).

\section{Spectral Type and Near-Infrared Colors}

Figure 4 shows plots of SDSS and MKO colors versus spectral type for the L and T dwarfs listed in Table~5 
of this paper and in Table~9 of K04.  Most of the spectral types have been assigned by us from our
own spectra, but a few types are adopted from other published work. (See notes for Table~5 and K04's 
Table~9.)  While the colors are generally correlated with spectral type, they are also significantly 
scattered.  The $J$--$H$ colors of L3--T1 dwarfs show a dispersion of $\sim$ 0.5~mag, as do the $H$--$K$ 
colors of all dwarfs later than type L3.  These dispersions are likely due to wide ranges of grain 
sedimentation properties and metallicity in the L dwarfs and wide ranges of gravity and metallicity in 
the T dwarfs (K04, and references therein).

As noted by K04, a small population of L dwarfs has unusually blue near-infrared colors, and we have found
in this sample a small 
population of T dwarfs with unusually red colors.  These unusual dwarfs are marked in Figure 4.  
Figure~5 shows the spectra of the anomalously blue mid-L 
dwarfs SDSS~J1033+4005, SDSS~J1422+2215 and SDSS~J1121+4332, along with the spectrum of the normal L6 
dwarf SDSS~J0654+6528.   The spectra of the blue dwarfs show strong H$_2$O and FeH absorption bands, 
which may be due to subsolar metallicity and/or thinner condensate cloud decks.  Figure~6 shows the
spectra of the unusually red early-T dwarfs SDSS~J1516+3053 and SDSS~J1415+5724, along with two 
normal reference spectra.  SDSS~J1516+3053 shows weak CH$_4$ bands but strong H$_2$O bands.  
SDSS~J1415+5724 shows an unusually weak $K$-band CH$_4$ band.  Attempts to reproduce these spectra as 
composite binary systems failed, but, as previously discussed, we may not know the true spectra of
single L--T transition dwarfs.  \citet{liu06} show that one of our new T dwarfs, SDSS~J1534+1615, 
is a close T1-T2 and T5-T6 binary whose brighter $J$-band component is the fainter one in $H$ and $K$.
This near-infrared flux inversion suggests that the system straddles the ``early T hump'' in the absolute magnitude-color
diagrams ({\it e.g.} K04, Golimowski et al. 2004).  Clearly, high-resolution imaging of dwarfs with unusual
colors is warranted, as are parallax and velocity studies to determine their kinematic properties. 

\section{Summary}

Thanks to the great gains in detection efficiency provided by large digital sky surveys, observers
have reached the once-unimaginable position of routinely discovering nearby field brown dwarfs.  In this 
paper, we report the discovery of 71 L and T dwarfs identified during our ongoing search for 
high-redshift quasars and brown dwarfs in SDSS imaging data.  Using near-infrared photometry
and spectra obtained over the past two years at UKIRT and IRTF, we have classified these dwarfs
using the L classification scheme of \citet{g02} and the new unified T classification scheme of
\citet{burg05}.  The near-infrared colors of the 71 dwarfs exhibit the same trends, scatter, and 
anomalies noted in previously reported samples.  The unusual spectral features exhibited by some 
dwarfs may lead to a better understanding of their underlying atmospheric properties.  We plan
to apply our results to the revision of the spectral infrared indices used in the near-infrared
classification of L dwarfs.  We will also integrate the 71 new L and T into a complete 
magnitude-limited catalog for the purpose of determining the substellar luminosity function.

\section{Acknowledgements}

We thank the staff at APO, IRTF, and UKIRT for their assistance with the observations and
data acquisition.  Some data were obtained through the UKIRT Service Programme.   UKIRT is 
operated by the Joint Astronomy Centre on behalf of the U.K. Particle Physics and Astronomy 
Research  Council.  APO is owned and operated by the Astrophysical Research Consortium (ARC).  
The IRTF is operated by the University of Hawaii under Cooperative Agreement no. NCC 5-538 
with the National Aeronautics and Space Administration (NASA), Office of Space Science, Planetary 
Astronomy Program.  This publication makes use of data products from the Two Micron All Sky 
Survey, which is a joint project of the University of Massachusetts and the Infrared 
Processing and Analysis Center (IPAC)/California Institute of Technology, funded by NASA and 
the National Science Foundation.  This research has also made use of the NASA/IPAC Infrared 
Science Archive, which is operated by the Jet Propulsion Laboratory, California Institute of 
Technology, under contract with NASA.

XF acknowledges support from NSF grant AST 03-07384,
a Sloan Research Fellowship and a David and Lucile Packard Fellowship. TRG's research is 
supported by the Gemini Observatory, which is operated by the Association of Universities for 
Research in Astronomy on behalf of the international Gemini  partnership of Argentina, Australia, 
Brazil, Canada, Chile, the United Kingdom, and the United States of  America.  

The Sloan Digital Sky Survey (SDSS) is managed by ARC for the Participating Institutions: The 
University of Chicago, Fermilab, the Institute for Advanced Study, the Japan Participation Group, 
The Johns Hopkins University, the Korean Scientist Group, Los Alamos National Laboratory, the 
Max-Planck-Institute for Astronomy, the Max-Planck-Institute for Astrophysics, New Mexico State 
University, University of Pittsburgh, University of Portsmouth, Princeton University, the United 
States Naval Observatory, and the University of Washington.  Funding for SDSS has been provided 
by the Alfred P. Sloan Foundation, the Participating Institutions, the National Aeronautics and 
Space Administration, the National Science Foundation, the U.S. Department of Energy, the Japanese 
Monbukagakusho, and the Max Planck Society.  The SDSS Web site is http://www.sdss.org/.

\clearpage
\LongTables
\begin{landscape}
\begin{deluxetable}{lccccccccc}
\tabletypesize{\scriptsize}
\tablewidth{0pt}
\tablenum{1}
\tablecolumns{10}
\tablecaption{New M, L, and T Dwarf Photometry}
\tablehead
{
     & SDSS $i$\tablenotemark{a}   & SDSS $z$   & SDSS & UT Date 	  & MKO $J$ & MKO $H$  & MKO $K$ &            & UT Date \\
SDSS Name &  (AB)   &  (AB)   &  Run & (yyyymmdd)  &  (Vega)   & (Vega)    & (Vega) & Instrument & (yyyymmdd)
}
\startdata

J000250.98$+$245413.8 &  $22.20\pm0.16$	    	&	$19.99\pm0.08$   &   4851 &    20040923   &      $17.02\pm 0.03$  &  $16.09\pm 0.03$  &  $15.42\pm 0.03$  &   UFTI  &  20050727  \\
J003609.26$+$241343.3 &  [$22.97\pm0.39$]	    &	$20.07\pm0.11$   &   4836 &    20040917   &      $17.08\pm 0.05$  &  $16.28\pm 0.03$  &  $15.48\pm 0.03$  &   UFTI  &  20050727  \\
J011912.22$+$240331.6 &  [$24.07\pm0.49$]	    &	$20.46\pm0.11$   &   4829 &    20040915   &      $16.86\pm 0.03$  &  $16.46\pm 0.03$  &  $16.26\pm 0.03$  &   SpeX  &  20050811  \\
J020608.97$+$223559.2 &  $22.42\pm0.17$	    	&	$19.25\pm0.04$   &   4844 &    20040921	  &	 $16.40\pm 0.05$  &  $15.56\pm 0.05$  &  $15.03\pm 0.05$  &   UFTI  &  20050208  \\
J020608.97$+$223559.2 &  $22.42\pm0.17$	    	&	$19.25\pm0.04$   &   4844 &    20040921	  &	 $16.31\pm 0.03$  &  $15.62\pm 0.03$  &  $15.05\pm 0.03$  &   SpeX  &  20050812  \\
& \\
J024256.98$+$212319.6 &  $22.24\pm 0.18$           &    $20.02\pm 0.09$   &   4844 &    20040921   &      $17.05\pm0.03$   &  $16.19\pm 0.03$  &  $15.43\pm 0.03$  &   SpeX  &  20051107 \\
J024749.90$-$163112.6 &  [$23.24\pm0.39$]       &   $19.83\pm0.11$   &   5069 &    20041215   &      $16.73\pm 0.03$  &  $16.31\pm 0.03$  &  $15.81\pm 0.03$  &   SpeX  &  20050812  \\
J032553.17$+$042540.1 &  [$23.59\pm0.57$]       &   $19.75\pm0.08$   &   5065 &    20041214   &      $15.88\pm 0.03$  &  $16.24\pm 0.03$  &  $16.48\pm 0.03$  &   SpeX  &  20050811  \\
J032553.17$+$042540.1 &  [$23.59\pm0.57$]       &   $19.75\pm0.08$   &   5065 &    20041214   &      $15.92\pm 0.03$  &  $16.26\pm 0.03$  &  $16.45\pm 0.05$  &   SpeX  &  20050812  \\
J035104.37$+$481046.8 &  [$22.84\pm0.30$]	    &	$19.57\pm0.06$   &   4887 &    20041015   &	 $16.33\pm 0.05$  &  $15.80\pm 0.05$  &  $15.15\pm 0.05$  &   UFTI  &  20050208  \\
& \\
J035104.37$+$481046.8 &  [$22.84\pm0.30$]	    &	$19.57\pm0.06$   &   4887 &    20041015   &	 $16.27\pm 0.03$  &  $15.81\pm 0.03$  &  $15.20\pm 0.03$  &   SpeX  &  20050812  \\
J065405.63$+$652805.4 &  $20.97\pm0.05$	   	 &	$18.89\pm0.04$   &   5060 &    20041213   &	 $16.08\pm 0.03$  &  $15.35\pm 0.03$  &  $14.59\pm 0.03$  &   SpeX  &  20050121  \\
J073922.26$+$661503.5 &  [$23.90\pm0.33$]	    &	$19.86\pm0.08$   &   4887 &    20041015   &	 $16.75\pm 0.03$  &  $16.31\pm 0.03$  &  $16.05\pm 0.03$  &   SpeX  &  20050407  \\
J082030.12$+$103737.0 &  [$22.75\pm0.25$]	    &	$20.03\pm0.09$   &   5194 &    20050312   &	 $17.03\pm 0.03$  &  $16.09\pm 0.05$  &  $15.58\pm 0.03$  &   SpeX  &  20050407  \\
J083506.16$+$195304.4 &  $21.10\pm0.07$	    	&	$18.84\pm0.05$   &   5045 &    20041212   &	 $15.94\pm 0.03$  &  $15.12\pm 0.03$  &  $14.41\pm 0.03$  &   SpeX  &  20050121  \\
& \\
J085116.20$+$181730.0 &  $21.92\pm0.15$	    	&	$19.68\pm0.08$   &   5061 &    20041213   &	 $16.68\pm 0.03$  &  $15.80\pm 0.04$  &  $14.97\pm 0.03$  &   SpeX  &  20050121  \\
J085834.42$+$325627.7 &  $22.09\pm0.15$	    	&	$19.07\pm0.05$   &   3606 &    20030125   &	 $16.33\pm 0.03$  &  $15.45\pm 0.03$  &  $14.72\pm 0.03$  &   UFTI  &  20040114  \\
J090900.73$+$652527.2 &  [$23.46\pm0.38$]	    &	$18.75\pm0.05$   &   4264 &    20031120   &	 $15.81\pm 0.03$  &  $15.32\pm 0.03$  &  $15.19\pm 0.03$  &   SpeX  &  20050121  \\
J100711.74$+$193056.2 &  [$22.93\pm0.29$]	    &	$19.76\pm0.09$   &   5183 &    20050310   &	 $16.75\pm 0.05$  &  $15.85\pm 0.05$  &  $15.15\pm 0.03$  &   SpeX  &  20050407  \\
J102552.43$+$321234.0 &  [$22.67\pm0.30$]	    &	$19.69\pm0.08$   &   4576 &    20040416   &	 $16.89\pm 0.05$  &  $15.98\pm 0.03$  &  $15.16\pm 0.03$  &   UFTI  &  20040626  \\
& \\
J102751.48$+$400931.7\tablenotemark{b}&$22.14\pm 0.16$ &  $19.98\pm 0.08$ & 3818 & 20030326 &      $17.10 \pm 0.05$  & $16.49 \pm  0.03$ & $15.87 \pm  0.03$ &   UFTI  &  20040118  \\
J103321.92$+$400549.5 &  $21.94\pm0.16$	   	 &	$19.54\pm0.06$   &   3647 &    20030201   &	 $16.74\pm 0.03$  &  $16.05\pm 0.04$  &  $15.60\pm 0.05$  &   SpeX  &  20050121  \\
J103931.35$+$325625.5 &  [$22.78\pm0.31$]	    &	$19.41\pm0.06$   &   4576 &    20040416   &	 $16.16\pm 0.03$  &  $15.47\pm 0.03$  &  $15.01\pm 0.03$  &   UFTI  &  20041229  \\
J104335.08$+$121314.1 &  $21.99\pm0.19$	    	&	$18.86\pm0.04$   &   3836 &    20030331   &	 $15.82\pm 0.03$  &  $14.87\pm 0.03$  &  $14.20\pm 0.03$  &   UFTI  &  20040118  \\
J104829.21$+$091937.8 &  [$24.31\pm0.60$]	    &	$19.69\pm0.08$   &   3031 &    20020312   &	 $16.39\pm 0.03$  &  $15.95\pm 0.03$  &  $15.87\pm 0.03$  &   UFTI  &  20040118  \\
& \\
J105213.51$+$442255.7 &  [$22.94\pm0.40$]	    &	$19.11\pm0.06$   &   3530 &    20021213   &	 $15.89\pm 0.03$  &  $15.09\pm 0.03$  &  $14.46\pm 0.03$  &   UFTI  &  20040118  \\
J111320.16$+$343057.9 &  $21.99\pm0.14$	    	&	$19.76\pm0.07$   &   4550 &    20040413   &	 $16.92\pm 0.03$  &  $16.00\pm 0.03$  &  $15.26\pm 0.03$  &   UFTI  &  20040626  \\
J112118.57$+$433246.5 &  $22.31\pm0.19$	    	&	$19.93\pm0.09$   &   3813 &    20030324   &	 $17.04\pm 0.05$  &  $16.56\pm 0.04$  &  $16.13\pm 0.04$  &   SpeX  &  20050121  \\
J114220.63$+$114440.3\tablenotemark{b}&$22.46\pm 0.25$ &  $20.24\pm 0.11$ & 3836 & 20030331 &      $17.89 \pm 0.06$  & $17.40 \pm  0.05$ & $16.84 \pm  0.09$ &   UFTI  &  20040626  \\
J120602.51$+$281328.7 &  [$24.11\pm0.85$]	    &	$19.48\pm0.07$   &   5112 &    20050117   &	 $16.10\pm 0.03$  &  $15.83\pm 0.03$  &  $15.97\pm 0.03$  &   UFTI  &  20050305  \\
& \\
J121440.95$+$631643.4 &  [$24.90\pm0.78$]	    &	$19.65\pm0.10$   &   2304 &    20010518   &	 $16.05\pm 0.03$  &  $15.80\pm 0.03$  &  $15.73\pm 0.04$  &   SpeX  &  20050121  \\
J121659.17$+$300306.3 &  $22.41\pm0.24$	    	&	$19.71\pm0.10$   &   5061 &    20041213   &	 $16.89\pm 0.05$  &  $16.19\pm 0.03$  &  $15.56\pm 0.03$  &   SpeX  &  20050407  \\
J121951.45$+$312849.4 &  $21.97\pm0.15$	    	&	$18.85\pm0.04$   &   4599 &    20040425   &	 $15.85\pm 0.03$  &  $14.98\pm 0.03$  &  $14.30\pm 0.03$  &   UFTI  &  20040626  \\
J123330.63$+$423948.8\tablenotemark{c}&[$22.79\pm 0.27$] &  $19.92\pm 0.09$ & 3840 & 20030401 &      $17.24 \pm 0.03$  & $16.33 \pm  0.03$ & $15.65 \pm  0.03$ &   UFTI  &  20040118  \\
J125011.65$+$392553.9 &  [$23.78\pm0.48$]	    &	$19.95\pm0.09$   &   3900 &    20030426   &	 $16.12\pm 0.03$  &  $16.07\pm 0.03$  &  $16.07\pm 0.03$  &   UFTI  &  20040118  \\
& \\
J134203.11$+$134022.2 &  $22.31\pm0.20$	    	&	$19.86\pm0.09$   &   3971 &    20030530   &	 $16.90\pm 0.03$  &  $15.95\pm 0.03$  &  $15.13\pm 0.03$  &   UFTI  &  20040626  \\
J134403.84$+$083950.9\tablenotemark{d}&[$23.09\pm 0.31$] &  $19.98\pm 0.08$ & 3909 & 20030428 &      $17.22 \pm 0.03$  & $16.45 \pm  0.03$ & $15.98 \pm  0.03$ &   UFTI  &  20040120  \\
J134525.57$+$521634.0 &  $22.02\pm0.17$	    	&	$19.84\pm0.10$   &   3177 &    20020508   &	 $17.05\pm 0.03$  &  $16.30\pm 0.03$  &  $15.54\pm 0.03$  &   SpeX  &  20050407  \\
J135852.68$+$374711.9 &  [$25.34\pm0.45$]	    &	$19.89\pm0.08$   &   3900 &    20030426   &	 $16.17\pm 0.03$  &  $16.41\pm 0.03$  &  $16.66\pm 0.05$  &   UFTI  &  20040118  \\
J140023.12$+$433822.3 &  [$22.52\pm0.20$]	    &	$19.08\pm0.04$   &   3716 &    20030311   &	 $16.16\pm 0.03$  &  $15.18\pm 0.03$  &  $14.47\pm 0.03$  &   UFTI  &  20040118  \\
& \\
J140255.66$+$080055.2 &  [$22.87\pm0.31$]	    &	$19.93\pm0.08$   &   3903 &    20030427   &	 $16.85\pm 0.03$  &  $16.25\pm 0.03$  &  $15.73\pm 0.03$  &   UFTI  &  20040120  \\
J141530.05$+$572428.7 &  [$22.56\pm0.30$]	    &	$19.87\pm0.09$   &   3225 &    20020609   &	 $16.49\pm 0.03$  &  $16.09\pm 0.03$  &  $15.55\pm 0.03$  &   UFTI  &  20040210  \\
J141659.78$+$500626.4 &  $22.18\pm0.23$	  	 &	$19.70\pm0.10$   &   3177 &    20020508   &	 $16.81\pm 0.03$  &  $16.04\pm 0.03$  &  $15.35\pm 0.03$  &   UFTI  &  20040214  \\
J141659.78$+$500626.4 &  $22.18\pm0.23$	    	&	$19.70\pm0.10$   &   3177 &    20020508   &	 $16.76\pm 0.05$  &  $16.01\pm 0.03$  &  $15.35\pm 0.03$  &   SpeX  &  20050407  \\
J142227.25$+$221557.1 &  $22.16\pm0.14$	    	&	$19.71\pm0.09$   &   4678 &    20040614   &	 $16.87\pm 0.03$  &  $16.12\pm 0.03$  &  $15.64\pm 0.03$  &   UFTI  &  20050306  \\
& \\
J143553.25$+$112948.6 &  [$23.40\pm0.44$]	    &	$20.13\pm0.10$   &   3996 &    20030622   &	 $17.04\pm 0.03$  &  $16.52\pm 0.04$  &  $16.15\pm 0.06$  &   UFTI  &  20040626  \\
J143945.86$+$304220.6 &  [$23.64\pm0.49$]	    &	$20.24\pm0.10$   &   4570 &    20040415   &	 $16.97\pm 0.03$  &  $16.53\pm 0.03$  &  $16.34\pm 0.05$  &   UFTI  &  20040626  \\
J144128.52$+$504600.4 &  $21.83\pm0.15$	    	&	$19.87\pm0.08$   &   3180 &    20020508   &	 $16.86\pm 0.03$  &  $15.98\pm 0.03$  &  $15.18\pm 0.03$  &   UFTI  &  20040214  \\
J150411.63$+$102718.4 &  [$24.66\pm0.78$]	    &	$20.34\pm0.14$   &   3894 &    20030425   &	 $16.49\pm 0.03$  &  $16.92\pm 0.03$  &  $17.02\pm 0.03$  &   UFTI  &  20040120  \\
J151114.66$+$060742.9 &  $21.88\pm0.19$	   	 &	$19.09\pm0.06$   &   2391 &    20010616   &	 $15.83\pm 0.03$  &  $15.16\pm 0.03$  &  $14.52\pm 0.03$  &   UFTI  &  20050306  \\
& \\
J151506.11$+$443648.3 &  $22.07\pm0.15$	    	&	$19.50\pm0.06$   &   3180 &    20020508   &	 $16.54\pm 0.03$  &  $15.63\pm 0.03$  &  $14.86\pm 0.03$  &   UFTI  &  20040211  \\
J151643.01$+$305344.4 &  [$23.18\pm0.35$]	    &	$19.99\pm0.11$   &   4002 &    20030623   &	 $16.79\pm 0.03$  &  $15.86\pm 0.03$  &  $15.12\pm 0.03$  &   UFTI  &  20040211  \\
J152039.82$+$354619.8 &  $21.86\pm0.12$	    	&	$18.31\pm0.03$   &   3818 &    20030326   &	 $15.47\pm 0.03$  &  $14.56\pm 0.03$  &  $14.02\pm 0.03$  &   UFTI  &  20040211  \\
J153417.05$+$161546.1AB &  [$24.17\pm0.61$]	    &	$20.20\pm0.12$   &   5194 &    20050312   &	 $16.62\pm 0.03$  &  $16.37\pm 0.03$  &  $16.06\pm 0.03$  &   UFTI  &  20050305  \\
J153453.33$+$121949.2 &  $20.48\pm0.05$       	&   	$18.18\pm0.03$   &   5317 &    20050511   &      $15.27\pm 0.03$  &  $14.42\pm 0.03$  &  $13.69\pm 0.03$  &   SpeX  &  20050811  \\
& \\
J154009.36$+$374230.3 &  [$22.67\pm 0.26$] 	    &   $19.03\pm 0.06$  &   3965 &    20030529   &      $16.33 \pm 0.03$  & $15.42 \pm  0.03$ & $14.65 \pm  0.03$ &   UFTI  &  20040214  \\
J154508.93$+$355527.3 &  $22.28\pm0.18$	    	&	$19.87\pm0.09$   &   3965 &    20030529   &	 $16.97\pm 0.03$  &  $16.04\pm 0.03$  &  $15.29\pm 0.03$  &   UFTI  &  20040214  \\
J154849.02$+$172235.4 &  $21.32\pm0.08$	    &	$18.82\pm0.04$   &   4674 &    20040613   &	 $16.09\pm 0.03$  &  $15.24\pm 0.03$  &  $14.49\pm 0.03$  &   UFTI  &  20050305  \\
J161731.65$+$401859.7 &  $22.09\pm0.16$	    	&	$19.81\pm0.10$   &   3226 &    20020609   &	 $16.83\pm 0.03$  &  $15.69\pm 0.03$  &  $14.84\pm 0.03$  &   SpeX  &  20050407  \\
J162051.17$+$323732.1 &  [$23.44\pm0.48$]	    &	$20.01\pm0.11$   &   3964 &    20030529   &	 $17.17\pm 0.04$  &  $16.23\pm 0.03$  &  $15.40\pm 0.03$  &   UFTI  &  20040728  \\
& \\
J162255.27$+$115924.1 &  $21.93\pm0.17$	    	&	$19.43\pm0.06$   &   5194 &    20050312   &	 $16.89\pm 0.05$  &  $16.15\pm 0.03$  &  $15.46\pm 0.03$  &   SpeX  &  20050408  \\
J162429.36$+$125144.0\tablenotemark{d}&$20.79\pm 0.15$ &  $18.97\pm 0.05$ & 5323 & 20050512 &      $16.44 \pm 0.03$  & $15.72 \pm  0.03$ & $15.18 \pm  0.03$ &   UFTI  &  20050307  \\
J162838.77$+$230821.1 &  [$24.40\pm0.58$]	    &	$20.23\pm0.10$   &   3927 &    20030501   &	 $16.25\pm 0.03$  &  $16.63\pm 0.04$  &  $16.72\pm 0.03$  &   UFTI  &  20040728  \\
J163022.92$+$081822.0 &  [$23.82\pm0.61$]	    &	$20.13\pm0.10$   &   3996 &    20030622   &	 $16.18\pm 0.03$  &  $16.35\pm 0.03$  &  $16.41\pm 0.03$  &   UFTI  &  20040118  \\
J163359.23$-$064056.5 &  $21.31\pm0.08$       	&   	$18.76\pm0.04$   &   5384 &    20050605   &      $16.00\pm 0.03$  &  $15.25\pm 0.03$  &  $14.54\pm 0.03$  &   SpeX  &  20050811  \\
& \\
J163607.48$+$233601.6\tablenotemark{d}&$21.82\pm 0.19$ &  $19.57\pm 0.06$ & 3997 & 20030622 &      $16.86\pm 0.03$  &  $16.21\pm 0.03$  & $15.61\pm 0.03$   &   UFTI  & 20040728 \\
J164916.89$+$464340.0 &  $22.39\pm 0.19$      	&   	$19.73\pm0.09$   &   4011 &    20030625   &      $17.09 \pm 0.03$  & $16.33 \pm  0.03$ & $15.74 \pm  0.03$ &   UFTI  &  20040728  \\
J170005.43$+$154128.8\tablenotemark{d}&$21.53\pm 0.10$ &  $19.14\pm 0.05$ & 4014 & 20030626 &      $16.21 \pm 0.03$  & $15.65 \pm  0.03$ & $15.12 \pm  0.03$ &   UFTI  &  20040728  \\
J171147.17$+$233130.5 &  $21.90\pm0.24$       	&   	$19.95\pm0.13$   &   3177 &    20020508   &      $16.84\pm 0.03$  & $16.00\pm 0.03$   & $15.25\pm 0.03$   &   SpeX  &  20050813  \\
J171902.15$+$373453.6\tablenotemark{d}&$21.87\pm0.13$  &  $20.28\pm 0.11$ & 4679 & 20040614 &      $17.65\pm 0.05$  &  $17.08\pm 0.04$  &  $16.50\pm 0.03$  &   SpeX  &  20050811  \\
& \\
J173101.41$+$531047.9 &  $21.55\pm0.14$	    	&	$19.34\pm0.07$   &   1336 &    20000404   &	 $16.32\pm 0.03$  &  $15.50\pm 0.03$  &  $14.78\pm 0.03$  &   UFTI  &  20050307  \\
J204317.69$-$155103.4 &  [$23.74\pm0.71$]       &   $19.72\pm0.10$   &   5415 &    20050611   &      $16.87\pm 0.03$  &  $16.02\pm 0.03$  &  $15.41\pm 0.03$  &   UFTI  &  20050801 \\
J205235.31$-$160929.8 &  [$22.72\pm0.23$]       &   $19.06\pm0.05$   &   5421 &    20050622   &      $16.04\pm 0.04$  &  $15.45\pm 0.03$  &  $15.00\pm 0.03$  &   SpeX  &  20050811 \\
J213154.43$-$011939.3 &  [$22.68\pm0.31$]	    &	$20.01\pm0.10$   &   4822 &    20040912   &	$17.29\pm 0.03$  &  $16.45\pm 0.03$  &  $15.75\pm 0.03$  &   UFTI  &  20050801 \\
J213240.36$+$102949.4 &  $21.49\pm0.10$	   	 &	$19.25\pm0.05$   &   2566 &    20010918   &	$16.38\pm 0.03$  &  $15.52\pm 0.03$  &  $14.76\pm 0.03$  &   SpeX  &  20050813 \\
& \\
J213352.72$+$101841.0 &  $21.96\pm0.16$	    &	$19.76\pm0.10$   &   1739 &    20000927   &	$16.92\pm 0.03$  &  $16.19\pm 0.03$  &  $15.61\pm 0.03$  &   SpeX  &  20050813 \\
J232804.58$-$103845.7 &  $21.44\pm0.15$	    &	$19.76\pm0.12$   &   1891 &    20001127   &	$16.77\pm 0.03$  &  $15.99\pm 0.03$  &  $15.20\pm 0.03$  &   SpeX  &  20050814 \\
\enddata
\tablenotetext{a}{Bracketed values denote asinh magnitudes of dwarfs not detected in SDSS $i$.} 
\tablenotetext{b}{Photometrically identified M dwarf; no spectrum obtained.} 
\tablenotetext{c}{Photometrically identified mid-L dwarf; no spectrum obtained.}
\tablenotetext{d}{Photometrically identified early-L dwarf; no spectrum obtained.}

\end{deluxetable}
\clearpage
\end{landscape}

\begin{deluxetable}{lcccccc}
\tablewidth{0pt}
\tablenum{2}
\tablecolumns{7}
\tablecaption{Additional $ZJHK$ Photometry   }
\tablehead
{Name & UFTI $Z$ & MKO $J$ & MKO $H$  & MKO $K$   & Instrument & UT Date
}
\startdata

2MASS J00345157$+$0523050 & \nodata & $15.11 \pm 0.03$ & $15.55\pm 0.03$ & $15.96\pm  0.03$ &   UFTI   & 20040830\\
SDSS J075840.33$+$324723.4  & $16.47 \pm 0.05$ & \nodata & \nodata & \nodata &    UFTI  & 20040329 \\
SDSS J080531.80$+$481233.0  & $16.42 \pm 0.05$ & \nodata & \nodata & \nodata  &   UFTI  & 20040329 \\		    
SDSS J093109.56$+$032732.5  & $18.40 \pm 0.05$ & \nodata & \nodata & \nodata  &   UFTI  & 20040329 \\		    
SDSS J111010.01$+$011613.1  & $18.00 \pm 0.05$ & \nodata & \nodata & \nodata &   UFTI	& 20040222 \\
SDSS J115700.50$+$061105.2  & $18.82 \pm 0.05$ & \nodata & \nodata & \nodata  &   UFTI  & 20040222 \\
2MASS J12095613$-$1004008   & $17.52 \pm 0.05$ & $15.55\pm 0.03$ & $15.24\pm 0.03$ & $15.17\pm 0.03$  &   UFTI  & 20040329 \\
SDSS J133148.90$-$011651.4  & $17.04 \pm 0.05$ & \nodata & \nodata & \nodata  &   UFTI  & 20040222 \\
SDSS J152103.24$+$013142.7  & $17.99 \pm 0.05$ & \nodata & \nodata & \nodata  &   UFTI  & 20040222 \\
2MASS J21011544$+$1756586 & \nodata & $16.81\pm 0.03$ & $15.89\pm 0.03$ & $15.00\pm 0.03$ & SpeX & 20050814 \\

\enddata
\end{deluxetable}

\clearpage
\LongTables
\begin{landscape}
\begin{deluxetable}{lllllllllll}
\tabletypesize{\scriptsize}
\tablewidth{0pt}
\tablenum{3}
\tablecolumns{11}
\tablecaption{Spectral Indices of L--T1 Dwarfs (Geballe et al. 2002 scheme)}
\tablehead
{
Name & \multicolumn{2}{c}{H$_2$O-$J$} & \multicolumn{2}{c}{H$_2$O-$H$} & \multicolumn{2}{c}{CH$_4$-$H$} & \multicolumn{2}{c}{CH$_4$-$K$} & Mean  & Instrument and \\
       &  Index & Type &  Index & Type &  Index & Type &  Index & Type & Type\tablenotemark{a}   & UT Date (yymmdd)
}
\startdata

SDSS J000250.98$+$245413.8 &  \nodata 	& \nodata &    1.57   &  L5     &      0.97    &	$<$T0	 & \nodata & \nodata  &    L5 $\pm$ 1	        & 	CGS4 $H$ 050802 \\
SDSS J000250.98$+$245413.8 &  1.39     &  $<$T0  &    [2.17]   & [T0]    &      0.91    &        $<$T0	 &    1.06 &   L6     &    L6 $\pm$ 1            & 	SpeX  050813 \\
SDSS J003609.26$+$241343.3 &  1.37     &  $<$T0  &    1.59   &  L5     &      0.97    &         $<$T0	 &    1.06 &   L6     &    L5.5                 & 	SpeX  050811 \\
SDSS J020608.97$+$223559.2 &  1.40  	&  $<$T0  &    1.56   &  L5     &      0.92    &	$<$T0	 &    1.04 &   L6     &    L5.5 		& 	SpeX  050123 \\  
SDSS J024256.98$+$212319.6 &  1.31     & $<$T0    &    1.53   &  $<$T0  &      1.02    &         L4      &    0.92 &   L3.5   &    L4                   &       SpeX  051105 \\
          & \\
SDSS J035104.37$+$481046.8 &  2.14  	&  T2.5   &    2.81   &  T2     &      1.18    &	T1.5	 &    1.11 &   L7.5   &    T1 $\pm$ 2	    	& 	SpeX  050121 \\  
SDSS J065405.63$+$652805.4 &  1.29  	&  $<$T0  &   [1.28]  & [L0.5]  &      0.97    &	\nodata	 &    1.04 &   L6     &    L6  			&   	SpeX  050123			    \\ 
SDSS J074007.30$+$200921.9 &  \nodata 	& \nodata &    1.56   &  L5     &      0.99    &	$<$T0	 &    1.13 &   L7.5   &    L6 $\pm$ 1.5 	& 	CGS4  $H$ 040920, $K$ 040916 \\
SDSS J075656.54$+$231458.5 &  \nodata 	& \nodata &    1.44   &  L2.5   &      0.94    &	$<$T0	 &    0.98 &   L4.5   &    L3.5 $\pm$ 1 	& 	CGS4  $H$,$K$ 040916 \\
SDSS J080531.80$+$481233.0 &  1.49  	& $<$T0   & \nodata   & \nodata &      1.12    &	T1	 &    1.14 &   L8     &    L9.5 $\pm$ 1.5 	& 	SpeX  050122 \\
          & \\
SDSS J080959.01$+$443422.2 &  1.45  	&  $<$T0  &    1.63   &  L6     &      0.96    &	$<$T0	 &    1.02 &   L5.5   &    L6			& 	CGS4  $J$ 040119, UIST $HK$ 040422 \\
SDSS J082030.12$+$103737.0 &  1.46  	&  $<$T0  &    2.59   &  T1.5   &      0.99    &	$<$T0	 &    1.12 &   L7.5     &    L9.5 $\pm$ 2     	& 	SpeX  050407			    \\  
SDSS J083506.16$+$195304.4 &  1.38  	&  $<$T0  &    1.54   &  L4.5   &      0.97    &        $<$T0    &    0.99 &   L5     &    L4.5 		&   	SpeX  050123			    \\ 
SDSS J085116.20$+$181730.0 &  1.48  	&  $<$T0  &    1.63   &  L6     &      0.97    &	$<$T0	 &    0.91 &   $<$~L3 &    L4.5 $\pm$ 1.5   	& 	SpeX  050406			    \\  
SDSS J085834.42$+$325627.7 &  \nodata 	& \nodata &    2.06   &  T0     &      1.06    &    	T0.5     &    1.08 &   L6.5   &    L9 $\pm$ 1.5             & 	UIST  $HK$ 040409 \\
          & \\
SDSS J085834.42$+$325627.7 &  1.84  	&  T1     &    2.03   &  T0     &      1.05    &   	T0       &    1.07 &   L6.5   &    L9 $\pm$ 1.5            & 	SpeX  050123 \\
SDSS J100711.74$+$193056.2 &  1.59  	&  T0     &    1.74   &  L8     &      1.01    &	 $<$T0	 &    1.07 &   L6.5   &    L8 $\pm$ 1.5      	& 	SpeX  050406			    \\  
SDSS J102552.43$+$321234.0 &  \nodata 	& \nodata &    1.57   &  L5     &      1.05    &   	T0       & \nodata & \nodata  &    L7.5 $\pm$ 2.5   	&       CGS4  $H$ 041229 \\
SDSS J102552.43$+$321234.0 &  1.64  	&  T0     &    1.55   &  L4.5   &      0.99    &    	$<$T0    &    1.04 &   L5.5   &    L6.5 $\pm$ 2.5   	&  	SpeX  050123		\\
SDSS J103321.92$+$400549.5 &  1.41  	&  $<$T0  &    1.63   &  L6     &      0.89    &	 $<$T0	 &    1.02 &   L5.5   &    L6			&   	SpeX  050408			    \\ 
          & \\
SDSS J103931.35$+$325625.5 &  \nodata 	& \nodata &    2.15   &  T0.5   &      1.17    &    	T1.5     & \nodata & \nodata  &    T1                   &   	CGS4  $H$ 041229\\
SDSS J103931.35$+$325625.5 &  1.92  	&  T1.5   &    2.76   &  T2     &      1.07    &   	 T0.5    &    1.44 &   T1     &    T1    		& 	SpeX  050122 \\
SDSS J104335.08$+$121314.1 &  \nodata 	& \nodata &    1.63   &  L6     &      0.97    &	 $<$T0	 &    1.17 &   L8.5   &    L7 $\pm$ 1	    	& 	UIST  $HK$ 040406			    \\  
SDSS J105213.51$+$442255.7 &  1.86  	&  T1.5   &    1.87   &  L9     &      1.09    &	 T0.5	 &    1.26 &   L9.5   &    T0 $\pm$ 1	    	& 	CGS4  $J$ 040430, UIST $HK$ 040405	    \\  
SDSS J111320.16$+$343057.9 &  1.33  	&  $<$T0  &    1.47   &  L3     &      0.85    &   	$<$T0    &    0.86 &   $<$L3  &    L3	      		&   	SpeX  050408			    \\ 
          & \\
SDSS J112118.57$+$433246.5 &  1.42  	&  $<$T0  &    1.68   &  L7     &      0.96    &   	$<$T0    &    1.14 &   L8     &	   L7.5	      		&   	SpeX  050406			    \\ 
SDSS J121659.17$+$300306.3 &  1.34  	&  $<$T0  &    1.42   &  L2.5   &      0.98    &	 $<$T0	 &    0.99 &   L4.5   &    L3.5 $\pm$ 1 	&	SpeX  050408			    \\  
SDSS J121951.45$+$312849.4 &  \nodata 	& \nodata &    1.79   &  L8.5   &      0.96    &	 $<$T0	 &    1.15 &   L8     &    L8			&	UIST  $HK$ 040703			    \\ 
SDSS J134203.11$+$134022.2 &  1.21  	&  $<$T0  &    1.57   &  L5     &      0.97    &	 $<$T0	 &    1.05 &   L6     &    L5.5 		&	SpeX  050406				    \\ 
SDSS J134525.57$+$521634.0 &  1.37  	&  $<$T0  &    1.47   &  L3     &      0.94    &	 $<$T0	 &    0.94 &   L3.5   &    L3.5			&	SpeX  050408			    \\ 
          & \\
SDSS J140023.12$+$433822.3 &  \nodata 	& \nodata &    1.65   &  L6.5   &     [1.04]   &	[T0]	 &    1.13 &   L7.5   &    L7 $\pm$ 1		&	UIST  $HK$ 040409			    \\  
SDSS J140255.66$+$080055.2 &  \nodata 	& \nodata &    2.06   &  T0     &      1.25    &   	T2       &    1.40 &   T0.5   &    T1+/-1		&	UIST  $HK$ 040429 \\
SDSS J140255.87$+$080055.6 &  \nodata 	& \nodata &    2.27   &  T0.5   &      1.19    &   	T1.5     &    1.43 &   T0.5   &    T1			&       UIST  $HK$ 040511 \\
SDSS J141659.78$+$500626.4 &  1.40  	&  $<$T0  &    1.73   &  L8     &      1.01    &	 $<$T0	 &    0.99 &   L5     &    L6.5 $\pm$ 1.5	&	SpeX  050408			    \\  
SDSS J141659.78$+$500626.4 &  \nodata 	& \nodata &    1.50   &  L3.5   &      1.01    &	 $<$T0	 &    1.01 &   L5     &    L4.5 $\pm$ 1	&	UIST  $HK$ 040604			    \\ 
          & \\
SDSS J142227.25$+$221557.1 &  1.17  	&  $<$T0  &    1.55   &  L4.5   &      1.01    &	 $<$T0	 &    1.17 &   L8.5   &    L6.5 $\pm$ 2 	&	SpeX  050406			    \\  
SDSS J144128.52$+$504600.4 &  \nodata 	& \nodata &    1.46   &  L3     &      0.97    &	 $<$T0	 &    0.95 &   L3.5   &    L3			&	UIST  $HK$ 040426			    \\ 
SDSS J151114.66$+$060742.9 &  1.92  	&  T1.5   &    1.82   &  L8.5   &      1.21    &	 T2	 &    1.11 &   L7     &    T0 $\pm$ 2		&	SpeX  050406			    \\  
SDSS J151506.11$+$443648.3 &  \nodata 	& \nodata &    1.90   &  L9     &      1.01    &	 $<$T0	 &    1.05 &   L6     &    L7.5 $\pm$ 1.5	&	UIST  $HK$ 040428			    \\  
SDSS J151643.01$+$305344.4 &  \nodata 	& \nodata &    2.94   &  T2.5   &      0.97    &	 $<$T0	 &    1.28 &   L9.5   &    T1 $\pm$ 1.5 	&	UIST  $HK$ 040605			    \\  
          & \\
SDSS J152039.82$+$354619.8 &  1.38  	&  $<$T0  &    1.84   &  L9     &      1.02    &	 L9.5	 &    1.25 &   L9.5   &    L9.5 		&	CGS4  $J$ 040622, UIST $HK$ 040405	    \\ 
SDSS J153453.33$+$121949.2 &  1.38   	&  $<$T0  &    1.44   &  L2.5   &      1.00    &	 $<$T0	 &    1.02 &   L5.5   &    L4 $\pm$ 1.5          &       SpeX  050812 \\
SDSS J154009.36$+$374230.3 &  1.40  	&  $<$T0  &    2.48   &  T1.5   &      0.88    &         $<$T0	 &    1.09 &   L7     &    L9 $\pm$ 1.5           &       SpeX  050813 \\
SDSS J154508.93$+$355527.3 &  \nodata 	& \nodata &    1.70   &  L7.5   &     [1.03]   &	[T0]	 &    1.14 &   L8     &    L7.5			&	UIST  $HK$ 040617			    \\ 
SDSS J154849.02$+$172235.4 &  1.27  	&  $<$T0  &    1.60   &  L5.5   &      0.94    &	 $<$T0	 &    0.99 &   L5     &    L5			&	SpeX  050406			    \\ 
          & \\
SDSS J161731.65$+$401859.7 & [1.52]  	&  [L9.5] &    1.55   &  L4.5   &      0.98    &	 $<$T0	 &    0.94 &   L3.5   &    L4			&	SpeX  050407			    \\ 
SDSS J162051.17$+$323732.1 &  \nodata 	& \nodata &   \nodata & \nodata &     \nodata  & 	\nodata  &    1.06 &   L6     &    L6			&	CGS4  $K$ 040916 			    \\ 
SDSS J162255.27$+$115924.1 &  1.41  	&  $<$T0  &    1.69   &  L7.5   &      0.90    &	 $<$T0	 &    0.99 &   L5     &    L6 $\pm$ 1.5 	&	SpeX  050408			    \\  
SDSS J163359.23$-$064056.5 &  1.40  	&  $<$T0  &    [2.03]   &  [T0]   &      0.88    &         L6      &    1.04 &   L6     &    L6	         	&	SpeX  050812 \\
SDSS J164916.89$+$464340.0 &  1.44     &  $<$T0  &    1.55   &  L4.5   &      0.99    &         $<$T0	 &    1.04 &   L6     &    L5			&	SpeX  050813 \\
          & \\
SDSS J171147.17$+$233130.5 &  1.06     &  $<$T0  &    1.39   &  L2     &      [1.04]    &         [T0]    &    1.02 &   L5.5   &    L3.5 $\pm$ 1.5	&	SpeX  050814 \\
SDSS J173101.41$+$531047.9 &  1.50  	&  $<$T0  &    1.70   &  L7.5   &      0.98    &	 $<$T0	 &    0.98 &   L4.5   &    L6 $\pm$ 1.5 	&	SpeX  050407			    \\  
SDSS J204317.69$-$155103.4 &  \nodata 	& \nodata &    1.82   &  L8.5   &      1.01    &	 $<$T0	 & \nodata & \nodata  &    L8.5 $\pm$ 1 	&       CGS4  $H$ 050802 \\
SDSS J204317.69$-$155103.4 &  1.53     &  L9.5   &    1.74   &  L8     &      1.06    &         T0      &    1.29 &   T0     &    L9.5 $\pm$ 1 	&       SpeX  050812 \\
SDSS J205235.31$-$160929.8 &  1.58     &  T0     &    1.84   &  L9     &      1.07    &         T0.5    &    1.28 &   L9.5   &    L9.5	                &       SpeX  050812 \\
          & \\
2MASS J21011544$+$1756586 &  1.30 & $<$T0 &    [2.10]   &  [T0]   &      0.92    &         $<$T0	 &    1.06 &   L6.5   &    L6.5 $\pm$ 1 	&       SpeX  050814 \\
SDSS J213154.43$-$011939.3 &  \nodata 	& \nodata &    1.95   &  L9.5   &      0.88    &	 $<$T0	 & \nodata & \nodata  &    L9.5 $\pm$ 1 	&       CGS4  $H$ 050801 \\
SDSS J213154.43$-$011939.3 &  1.50     &  $<$T0  &    1.88   &  L9     &      0.94    &         $<$T0	 &    1.14 &   L8     &    L8.5                 &       SpeX  050811 \\
SDSS J213240.36$+$102949.4 &  1.46     &  $<$T0  &    1.61   &  L5.5   &      0.91    &         $<$T0	 &    0.95 &   L3.5   &    L4.5 $\pm$ 1           &       SpeX  050813 \\
SDSS J213352.72$+$101841.0 &  1.30     &  $<$T0  &    1.52   &  L4     &      0.97    &         $<$T0	 &    1.04 &   L6     &    L5 $\pm$ 1             &       SpeX  050813 \\
          & \\
SDSS J232804.58$-$103845.7 &  1.33     &  $<$T0  &    1.50   &  L3.5   &      0.97    &         $<$T0	 &    0.93 &   L3.5   &    L3.5                 &       SpeX  050814 \\
\enddata
\tablenotetext{a}{Uncertainties in mean types are $\pm 0.5$ subtype, unless otherwise noted.}
\end{deluxetable}
\clearpage
\end{landscape}

\clearpage
\LongTables
\begin{landscape}
\begin{deluxetable}{lllllllllllll}
\tabletypesize{\scriptsize}
\tablewidth{0pt}
\tablenum{4}
\tablecolumns{13}
\tablecaption{Spectral Indices of L8--T Dwarfs (Burgasser et al. 2005 scheme)}
\tablehead
{
Name & \multicolumn{2}{c}{H$_2$O-$J$} & \multicolumn{2}{c}{CH$_4$-$J$} & \multicolumn{2}{c}{H$_2$O-$H$} & \multicolumn{2}{c}{CH$_4$-$H$} & \multicolumn{2}{c}{CH$_4$-$K$} & Mean  & Instrument and \\
       &  Index & Type &  Index & Type &  Index & Type &  Index & Type &  Index & Type & Type\tablenotemark{a}   & UT Date (yymmdd)
}
\startdata
SDSS J011912.22$+$240331.6 & 0.42 & T3    & 0.59 & T2.5 & 0.47 & T2.5 & 0.93 & T1.5 & 0.63 & T1.5  & T2           & SpeX 050812\\
SDSS J024749.90$-$163112.6 & 0.41 & T3    & 0.44 & T4.5 & 0.63 & T0   & 0.85 & T2   & 0.72 & T1    & T2 $\pm$ 1.5 & SpeX 050814\\
SDSS J032553.17$+$042540.1 & 0.20 & T5.5  & 0.39 & T5   & 0.28 & T6   & 0.35 & T5.5 & 0.20 & T5    & T5.5         & SpeX 050813\\
SDSS J035104.37$+$481046.8 & 0.48 & T2    & 0.64 & T2   & 0.56 & T1   & 0.85 & T2   & 0.90 & $<$T0 & T1 $\pm$ 1   & SpeX 050121\\
SDSS J073922.26$+$661503.5 & 0.61 & $<$T2 & 0.75 & T0   & 0.46 & T2.5 & 0.85 & T2   & 0.59 & T2    & T1.5 $\pm$ 1 & SpeX 050406\\
 & \\
SDSS J080531.80$+$481233.0 & 0.70    & $<$T2   & 0.78    & $\leq$T0 & 0.61 & T0.5 & 0.90 & T1.5  & 0.88 & $<$T0 & T0.5 $\pm$ 1 & SpeX 050122\\
SDSS J082030.12$+$103737.0 & 0.61    & $<$T2   & 0.70    & T1       & 0.63 & T0   & 1.01 & $<$T0 & 0.89 & $<$T0 & $<$T0        & SpeX 050407\\
SDSS J085834.42$+$325627.7 & \nodata & \nodata & \nodata & \nodata  & 0.60 & T0.5 & 0.94 & T1    & 0.93 & $<$T0 & T0.5         & UIST $HK$ 040409\\
SDSS J085834.42$+$325627.7 & 0.55    & T1.5    & 0.69    & T1       & 0.55 & T1   & 0.95 & T1    & 0.94 & $<$T0 & T1           & SpeX 050123\\
SDSS J090900.73$+$652527.2 & 0.54    & T1.5    & 0.71    & T1       & 0.48 & T2   & 0.84 & T2    & 0.59 & T2    & T1.5         & SpeX 050123\\
 & \\
SDSS J100711.74$+$193056.2 & 0.65    & $<$T2   & 0.80    & $<$T0   & 0.60 & L9.5  & 0.99 & T0    & 0.94    & $<$T0   & $<$T0 & SpeX 050406\\
SDSS J103931.35$+$325625.5 & \nodata & \nodata & \nodata & \nodata & 0.61 & T0.5  & 0.85 & T2    & \nodata & \nodata & T1    & CGS4 $H$ 041229\\
SDSS J103931.35$+$325625.5 & 0.54    & T1.5    & 0.66    & T1.5    & 0.59 & T0.5  & 0.94 & T1    & 0.69    & T1      & T1    & SpeX 050122\\ 
SDSS J104829.21$+$091937.8 & 0.47    & T2.5    & 0.48    & T4      & 0.51 & T2    & 0.83 & T2.5 & 0.55    & T2.5    &  T2.5 $\pm$ 1 & CGS4 $J$ 040514, UIST $HK$ 040409\\
SDSS J105213.51$+$442255.7 & 0.57    & $<$T2   & 0.60    & T2      & 0.68 & $<$T0 & 0.92 & T1.5 & 0.80    & T0      &  T0.5 $\pm$ 1 & CGS4 $J$ 040430, UIST $HK$ 040405\\
 & \\
SDSS J120602.51$+$281328.7 & \nodata & \nodata & \nodata & \nodata & 0.51 & T2    & 0.61 & T3.5 & \nodata & \nodata & T3            & CGS4 $H$ 050306\\
SDSS J120602.51$+$281328.7 & 0.45    & T2.5    & 0.55    & T3      & 0.46 & T2.5  & 0.71 & T3   & 0.47    & T3      &  T2.5         & SpeX 050408\\
2MASS J12095613$-$1004008  & 0.44    & T2.5    & 0.47    & T4      & 0.46 & T2.5  & 0.67 & T3   & 0.49    & T3      & T3            & CGS4 $Z$ 040627, $J$ 040430, UIST $HK$ 040430\\
SDSS J121440.95$+$631643.4 & 0.34    & T4      & 0.63    & T2      & 0.35    & T5      & 0.71    & T3      & 0.38    & T3      & T3.5 $\pm$ 1  & SpeX 050123\\     
SDSS J121951.45$+$312849.4 & \nodata & \nodata & \nodata & \nodata & 0.61    & T0.5    & 1.04    & $<$T0   & 0.87    & $<$T0   & $<$T0         & UIST $HK$ 040703\\
 & \\
SDSS J125011.65$+$392553.9 & 0.36    & T4      & 0.47    & T4      & 0.39    & T4      & 0.53    & T4      & 0.23    & T4.5    & T4            & CGS4 $J$ 040513, UIST $HK$ 040408\\
SDSS J135852.68$+$374711.9 & 0.30    & T4.5    & 0.43    & T4.5    & \nodata & \nodata & \nodata & \nodata & \nodata & \nodata & T4.5          & CGS4 $J$ 040531\\
SDSS J135852.68$+$374711.9 & 0.31    & T4.5    & 0.63    & T2      & 0.36    & T4.5    & 0.36    & T5.5    & 0.29    & T4      & T4 $\pm$ 1    & SpeX 050408\\
SDSS J140255.66$+$080055.2  & \nodata & \nodata& \nodata & \nodata& 0.54 & T1.5 & 0.80 & T2.5 & 0.72 & T1  &    T1.5 & UIST $HK$ 040429\\  
SDSS J140255.66$+$080055.2  & \nodata & \nodata& \nodata & \nodata& 0.57 & T1 & 0.84 & T2 & 0.70 & T1  &    T1.5 & UIST $HK$ 040511\\  
 & \\
SDSS J141530.05$+$572428.7 & 0.37 & T3.5 & 0.48 & T4 & 0.48 & T2 & 0.63 & T3.5 & 0.66 & T1.5 &  T3 $\pm$ 1  &       CGS4 $J$ 040609, UIST $HK$ 040430\\
SDSS J143553.25$+$112948.6  & \nodata & \nodata& \nodata & \nodata&   0.55 & T1 & 0.88 & T2 &  \nodata & \nodata&      T1.5 &       CGS4 $H$ 041229 \\
SDSS J143553.25$+$112948.6  & 0.41 & T3 & 0.56 & T3 & 0.49 & T2 & 0.75 & T3 & 0.70 & T1 &  T2.5 $\pm$ 1 & SpeX 050407\\
SDSS J143945.86$+$304220.6  & \nodata & \nodata& \nodata & \nodata&   0.57 & T1 & 0.70 & T3 &   \nodata & \nodata&      T2 $\pm$ 1 &       CGS4 $H$ 041229\\
SDSS J143945.86$+$304220.6  & 0.45 & T2.5 & 0.59 & T2.5 & 0.49 & T2 & 0.72 & T3 & 0.53 & T2.5 & T2.5 &       SpeX 050123\\
 & \\
SDSS J150411.63$+$102718.4  & 0.08 & T7.5 & 0.26 & T7 & 0.24 & T7 & 0.19 & T7 & 0.13 & T6.5  &       T7 &       CGS4 $J$ 040613, UIST $HK$ 040615 \\
SDSS J151114.66$+$060742.9  & 0.67 &  $<$T2 & 0.84 & $<$T0 &  0.67 &  $<$T0 &  0.83 & T2 & 0.90 & $<$T0 & T0? &       SpeX 050406 \\
SDSS J151643.01$+$305344.4  & \nodata & \nodata& \nodata & \nodata& 0.49 & T2 &  1.03 & $<$T0 & 0.78 & T0   &  T0.5 $\pm$ 1 &       UIST $HK$ 040605 \\
SDSS J152039.82$+$354619.8  & 0.76 & $<$T2 & 0.68 & T1.5 & 0.69 & $<$T0 & 0.98 & T0.5 & 0.80 & T0  & T0 $\pm$ 1 &       CGS4 $J$ 040622, UIST $HK$ 040405 \\
SDSS J153417.05$+$161546.1AB  & 0.34 & T4   & 0.52 & T3.5 &  0.39 & T4    & 0.72 & T3   & 0.56 & T2.5  &    T3.5 &       SpeX 050406 \\
 & \\
SDSS J154009.36$+$374230.3  & 0.61 & $<$T2 &  0.59 & T2.5 & 0.76 & $<$T0 &  1.14 & $<$T0 &  0.92 & $<$T0 & \nodata &       SpeX 050813 \\
SDSS J162838.77$+$230821.1  & 0.07 & T7.5 & 0.22 & T7.5 &  0.28 & T6    & 0.19 & T7   & \nodata & \nodata&         T7 &       CGS4 $J$ 040729, $H$ 040916 \\
SDSS J163022.92$+$081822.0  & 0.29 & T4.5   & 0.35 & T5.5 &  0.32 & T5.5  & 0.37 & T5.5 & 0.17 & T5.5 &   T5.5 &       CGS4 $J$ 040502, UIST $HK$ 040429 \\
SDSS J204317.69$-$155103.4  & 0.69 & $<$T2 &  0.66 & T1.5  & 0.66 & $<$T0 &  0.95 &   T1 & 0.77 & T0 &   T0.5 $\pm$ 1 & SpeX 050812 \\
SDSS J205235.31$-$160929.8  & 0.67 & $<$T2 &  0.56 & T3 &  0.67  & $<$T0 & 0.93 & T1.5 & 0.78 & T0 &      T1 $\pm$ 1 &       SpeX 050812 \\
 & \\
SDSS J212413.89$+$010000.3  & 0.25 & T5   &  0.33  &  T6 &  0.42 &  T3.5  & 0.43 &  T5 & 0.20 & T5  &       T5 $\pm$ 1 &       SpeX 050811 \\
SDSS J213154.43$-$011939.3  & \nodata & \nodata& \nodata & \nodata & 0.67 & $<$T0 &  1.13  & $<$T0     & \nodata & \nodata & \nodata &       CGS4 $H$ 050801 \\        
SDSS J213154.43$-$011939.3  &0.68   &  $<$T2 &  0.71 & T1 & 0.58 &  T1 & 1.07 &  $<$T0  & 0.88 & $<$T0  & T0? &       SpeX 050811 \\
\enddata
\tablenotetext{a}{Uncertainties in mean types are $\pm 0.5$ subtype, unless otherwise noted.}
\end{deluxetable}
\end{landscape}
\begin{deluxetable}{lllcrrrc}
\hspace*{-0.5in}
\tabletypesize{\scriptsize}
\tablewidth{0pt}
\tablenum{5}
\tablecolumns{8}
\tablecaption{L and T Dwarf Colors versus Spectral Type}
\tablehead
{ Name & Type\tablenotemark{a} & $i$--$z$\tablenotemark{b} & $z$--$J$ & $J$--$H$ & $H$--$K$  & $J$--$K$  &  $J$\tablenotemark{c}  }
\startdata
SDSS J111320.16$+$343057.9  &  L3.0           &   $ 2.23\pm0.16$   &    $2.84\pm0.08$ &   $0.92\pm0.04$	&   $0.74\pm0.04$  &   $1.66\pm0.04$	& 16.92   \\ 
SDSS J144128.52$+$504600.4  &  L3.0           &   $ 1.96\pm0.17$   &    $3.01\pm0.09$ &   $0.88\pm0.04$	&   $0.80\pm0.04$  &   $1.68\pm0.04$	& 16.86   \\ 
SDSS J075656.54$+$231458.5  &  L3.5 $\pm$ 1.0 &   $>2.68       $   &    $3.02\pm0.07$ &   $0.98\pm0.04$	&   $0.82\pm0.04$  &   $1.80\pm0.04$	& 16.80   \\ 
SDSS J121659.17$+$300306.3  &  L3.5 $\pm$ 1.0 &   $ 2.70\pm0.26$   &    $2.82\pm0.11$ &   $0.70\pm0.06$	&   $0.63\pm0.04$  &   $1.33\pm0.06$	& 16.89   \\ 
SDSS J134525.57$+$521634.0  &  L3.5           &   $ 2.18\pm0.20$   &    $2.79\pm0.10$ &   $0.75\pm0.04$	&   $0.76\pm0.04$  &   $1.51\pm0.04$	& 17.05   \\ 
 & \\
SDSS J171147.17$+$233130.5  &  L3.5 $\pm$ 1.5 &   $ 1.95\pm0.27$   &    $3.11\pm0.13$ &   $0.84\pm0.04$	&   $0.75\pm0.04$  &   $1.59\pm0.04$	& 16.84   \\ 
SDSS J232804.58$-$103845.7  &  L3.5           &   $ 1.68\pm0.19$   &    $2.99\pm0.12$ &   $0.78\pm0.04$	&   $0.79\pm0.04$  &   $1.57\pm0.04$	& 16.77   \\ 
SDSS J153453.33$+$121949.2  &  L4.0 $\pm$ 1.5 &   $ 2.30\pm0.06$   &    $2.91\pm0.04$ &   $0.85\pm0.04$	&   $0.73\pm0.04$  &   $1.58\pm0.04$	& 15.27   \\ 
SDSS J161731.65$+$401859.7  &  L4.0           &   $ 2.28\pm0.19$   &    $2.98\pm0.10$ &   $1.14\pm0.04$	&   $0.85\pm0.04$  &   $1.99\pm0.04$	& 16.83   \\ 
SDSS J024256.98$+$212319.6  &  L4.0           &   $ 2.22\pm0.20$   &    $2.97\pm0.09$ &   $0.86\pm0.04$   &   $0.76\pm0.04$  &   $1.62\pm0.04$    & 17.05   \\
 & \\
SDSS J083506.16$+$195304.4  &  L4.5           &   $ 2.26\pm0.09$   &    $2.90\pm0.06$ &   $0.82\pm0.04$	&   $0.71\pm0.04$  &   $1.53\pm0.04$	& 15.94   \\ 
SDSS J085116.20$+$181730.0  &  L4.5 $\pm$ 1.5 &   $ 2.24\pm0.17$   &    $3.00\pm0.09$ &   $0.88\pm0.05$	&   $0.83\pm0.05$  &   $1.71\pm0.04$	& 16.68   \\ 
SDSS J213240.36$+$102949.4  &  L4.5 $\pm$ 1.0 &   $ 2.24\pm0.11$   &    $2.87\pm0.06$ &   $0.86\pm0.04$	&   $0.76\pm0.04$  &   $1.62\pm0.04$	& 16.38   \\ 
SDSS J154849.02$+$172235.4  &  L5.0           &   $ 2.50\pm0.08$   &    $2.73\pm0.05$ &   $0.85\pm0.04$	&   $0.75\pm0.04$  &   $1.60\pm0.04$	& 16.09   \\ 
SDSS J164916.89$+$464340.0  &  L5.0           &   $ 2.66\pm0.21$   &    $2.64\pm0.09$ &   $0.76\pm0.04$	&   $0.59\pm0.04$  &   $1.35\pm0.04$	& 17.09   \\ 
 & \\
SDSS J213352.72$+$101841.0  &  L5.0 $\pm$ 1.0 &   $ 2.20\pm0.19$   &    $2.84\pm0.10$ &   $0.73\pm0.04$	&   $0.58\pm0.04$  &   $1.31\pm0.04$	& 16.92   \\ 
SDSS J000250.98$+$245413.8  &  L5.5           &   $ 2.21\pm0.18$   &    $2.97\pm0.09$ &   $0.93\pm0.04$	&   $0.67\pm0.04$  &   $1.60\pm0.04$	& 17.02   \\ 
SDSS J003609.26$+$241343.3  &  L5.5           &   $>2.43       $   &    $2.99\pm0.12$ &   $0.80\pm0.06$	&   $0.80\pm0.04$  &   $1.60\pm0.06$	& 17.08   \\ 
SDSS J020608.97$+$223559.2  &  L5.5           &   $ 3.17\pm0.17$   &    $2.91\pm0.05$ &   $0.75\pm0.05$	&   $0.56\pm0.05$  &   $1.30\pm0.05$	& 16.34   \\ 
SDSS J134203.11$+$134022.2  &  L5.5           &   $ 2.45\pm0.22$   &    $2.96\pm0.09$ &   $0.95\pm0.04$	&   $0.82\pm0.04$  &   $1.77\pm0.04$	& 16.90   \\ 
 & \\
SDSS J141659.78$+$500626.4  &  L5.5 $\pm$ 2.0 &   $ 2.48\pm0.25$   &    $2.91\pm0.10$ &   $0.76\pm0.05$	&   $0.68\pm0.04$  &   $1.44\pm0.05$	& 16.79   \\ 
SDSS J173101.41$+$531047.9  &  L6.0 $\pm$ 1.5 &   $ 2.21\pm0.16$   &    $3.02\pm0.08$ &   $0.82\pm0.04$	&   $0.72\pm0.04$  &   $1.54\pm0.04$	& 16.32   \\ 
SDSS J065405.63$+$652805.4  &  L6.0 $\pm$ 1.0 &   $ 2.08\pm0.06$   &    $2.81\pm0.05$ &   $0.73\pm0.04$	&   $0.76\pm0.04$  &   $1.49\pm0.04$	& 16.08   \\ 
SDSS J074007.30$+$200921.9  &  L6.0 $\pm$ 1.5 &   $>2.72       $   &    $3.11\pm0.10$ &   $0.85\pm0.04$	&   $0.71\pm0.04$  &   $1.56\pm0.04$	& 16.67   \\ 
SDSS J080959.01$+$443422.2  &  L6.0           &   $ 2.54\pm0.17$   &    $2.91\pm0.07$ &   $1.12\pm0.04$	&   $0.94\pm0.04$  &   $2.06\pm0.04$	& 16.37   \\ 
 & \\
SDSS J103321.92$+$400549.5  &  L6.0           &   $ 2.40\pm0.17$   &    $2.80\pm0.07$ &   $0.69\pm0.05$	&   $0.45\pm0.06$  &   $1.14\pm0.06$	& 16.74   \\ 
SDSS J162051.17$+$323732.1  &  L6.0           &   $>2.49       $   &    $2.84\pm0.12$ &   $0.94\pm0.05$	&   $0.83\pm0.04$  &   $1.77\pm0.05$	& 17.17   \\ 
SDSS J162255.27$+$115924.1  &  L6.0 $\pm$ 1.5 &   $ 2.50\pm0.18$   &    $2.54\pm0.08$ &   $0.74\pm0.06$	&   $0.69\pm0.04$  &   $1.43\pm0.06$	& 16.89   \\ 
SDSS J163359.23$-$064056.5  &  L6.0           &   $ 2.55\pm0.09$   &    $2.76\pm0.05$ &   $0.75\pm0.04$	&   $0.71\pm0.04$  &   $1.46\pm0.04$	& 16.00   \\ 
SDSS J142227.25$+$221557.1  &  L6.5 $\pm$ 2.0 &   $ 2.45\pm0.17$   &    $2.84\pm0.09$ &   $0.75\pm0.04$	&   $0.48\pm0.04$  &   $1.23\pm0.04$	& 16.87   \\ 
 & \\
2MASS J21011544$+$1756586   &  L6.5 $\pm$ 1.0 &   $ 2.76\pm0.28$   &    $2.87\pm0.09$ &   $0.92\pm0.04$	&   $0.89\pm0.04$  &   $1.81\pm0.04$	& 16.81   \\ 
SDSS J104335.08$+$121314.1  &  L7.0 $\pm$ 1.0 &   $ 3.13\pm0.19$   &    $3.04\pm0.05$ &   $0.95\pm0.04$	&   $0.67\pm0.04$  &   $1.62\pm0.04$	& 15.82   \\ 
SDSS J140023.12$+$433822.3  &  L7.0 $\pm$ 1.0 &   $>3.42       $   &    $2.92\pm0.05$ &   $0.98\pm0.04$	&   $0.71\pm0.04$  &   $1.69\pm0.04$	& 16.16   \\ 
SDSS J102552.43$+$321234.0  &  L7.5 $\pm$ 2.5 &   $>2.81       $   &    $2.80\pm0.09$ &   $0.91\pm0.06$	&   $0.82\pm0.04$  &   $1.73\pm0.06$	& 16.89   \\ 
SDSS J112118.57$+$433246.5  &  L7.5           &   $ 2.38\pm0.21$   &    $2.89\pm0.10$ &   $0.48\pm0.06$	&   $0.43\pm0.06$  &   $0.91\pm0.06$	& 17.04   \\ 
 & \\
SDSS J151506.11$+$443648.3  &  L7.5 $\pm$ 1.5 &   $ 2.57\pm0.16$   &    $2.96\pm0.07$ &   $0.91\pm0.04$	&   $0.77\pm0.04$  &   $1.68\pm0.04$	& 16.54   \\ 
SDSS J154508.93$+$355527.3  &  L7.5           &   $ 2.41\pm0.20$   &    $2.90\pm0.09$ &   $0.93\pm0.04$	&   $0.75\pm0.04$  &   $1.68\pm0.04$	& 16.97   \\ 
SDSS J100711.74$+$193056.2  &  L8.0 $\pm$ 1.5 &   $>2.74       $   &    $3.01\pm0.10$ &   $0.90\pm0.07$	&   $0.70\pm0.06$  &   $1.60\pm0.06$	& 16.75   \\ 
SDSS J121951.45$+$312849.4  &  L8.0           &   $ 3.12\pm0.16$   &    $3.00\pm0.05$ &   $0.87\pm0.04$	&   $0.68\pm0.04$  &   $1.55\pm0.04$	& 15.85   \\ 
SDSS J154009.36$+$374230.3  &  L9.0 $\pm$ 1.5 &   $>3.47       $   &    $2.70\pm0.07$ &   $0.91\pm0.04$	&   $0.77\pm0.04$  &   $1.68\pm0.04$	& 16.33   \\ 
 & \\
SDSS J204317.69$-$155103.4  &  L9.0           &   $>2.78       $   &    $2.85\pm0.10$ &   $0.85\pm0.04$	&   $0.61\pm0.04$  &   $1.46\pm0.04$	& 16.87   \\ 
SDSS J213154.43$-$011939.3  &  L9.0           &   $>2.49       $   &    $2.72\pm0.10$ &   $0.84\pm0.04$	&   $0.70\pm0.04$  &   $1.54\pm0.04$	& 17.29   \\ 
SDSS J080531.80$+$481233.0  &  L9.5 $\pm$ 1.5 &   $2.20\pm0.05 $   &    $3.01\pm0.04$ &   $0.60\pm0.04$	&   $0.50\pm0.04$  &   $1.10\pm0.04$	& 14.61   \\ 
SDSS J082030.12$+$103737.0  &  L9.5 $\pm$ 2.0 &   $>2.47       $   &    $3.00\pm0.09$ &   $0.94\pm0.06$	&   $0.51\pm0.06$  &   $1.45\pm0.04$	& 17.03   \\ 
SDSS J151114.66$+$060742.9  &  T0.0 $\pm$ 2.0 &   $ 2.79\pm0.20$   &    $3.26\pm0.07$ &   $0.67\pm0.04$	&   $0.64\pm0.04$  &   $1.31\pm0.04$	& 15.83   \\ 
 & \\
SDSS J152039.82$+$354619.8  &  T0.0 $\pm$ 1.0 &   $ 3.55\pm0.12$   &    $2.84\pm0.04$ &   $0.91\pm0.04$	&   $0.54\pm0.04$  &   $1.45\pm0.04$	& 15.47   \\ 
SDSS J105213.51$+$442255.7  &  T0.5 $\pm$ 1.0 &   $>3.39       $   &    $3.22\pm0.07$ &   $0.80\pm0.04$	&   $0.63\pm0.04$  &   $1.43\pm0.04$	& 15.89   \\ 
SDSS J151643.01$+$305344.4  &  T0.5 $\pm$ 1.0 &   $>2.51       $   &    $3.20\pm0.11$ &   $0.93\pm0.04$	&   $0.74\pm0.04$  &   $1.67\pm0.04$	& 16.79   \\ 
SDSS J035104.37$+$481046.8  &  T1.0 $\pm$ 1.5 &   $>2.93       $   &    $3.28\pm0.07$ &   $0.48\pm0.05$	&   $0.63\pm0.05$  &   $1.11\pm0.05$	& 16.29   \\ 
SDSS J085834.42$+$325627.7  &  T1.0           &   $ 3.02\pm0.16$   &    $2.74\pm0.06$ &   $0.88\pm0.04$	&   $0.73\pm0.04$  &   $1.61\pm0.04$	& 16.33   \\ 
 & \\
SDSS J103931.35$+$325625.5  &  T1.0           &   $>3.09       $   &    $3.25\pm0.07$ &   $0.69\pm0.04$	&   $0.46\pm0.04$  &   $1.15\pm0.04$	& 16.16   \\ 
SDSS J205235.31$-$160929.8  &  T1.0 $\pm$ 1.0 &   $>3.44       $   &    $3.02\pm0.06$ &   $0.59\pm0.05$	&   $0.45\pm0.04$  &   $1.04\pm0.05$	& 16.04   \\ 
SDSS J073922.26$+$661503.5  &  T1.5 $\pm$ 1.0 &   $>2.64       $   &    $3.11\pm0.09$ &   $0.44\pm0.04$	&   $0.26\pm0.04$  &   $0.70\pm0.04$	& 16.75   \\ 
SDSS J090900.73$+$652527.2  &  T1.5           &   $>3.75       $   &    $2.94\pm0.06$ &   $0.49\pm0.04$	&   $0.13\pm0.04$  &   $0.62\pm0.04$	& 15.81   \\ 
SDSS J140255.66$+$080055.2  &  T1.5           &   $>2.57       $   &    $3.08\pm0.09$ &   $0.60\pm0.04$	&   $0.52\pm0.04$  &   $1.12\pm0.04$	& 16.85   \\ 
 & \\
SDSS J011912.22$+$240331.6  &  T2.0           &   $>2.04       $   &    $3.60\pm0.11$ &   $0.40\pm0.04$	&   $0.20\pm0.04$  &   $0.60\pm0.04$	& 16.86   \\ 
SDSS J024749.90$-$163112.6  &  T2.0 $\pm$ 1.5 &   $>2.67       $   &    $3.10\pm0.11$ &   $0.42\pm0.04$	&   $0.50\pm0.04$  &   $0.92\pm0.04$	& 16.73   \\ 
SDSS J143553.25$+$112948.6  &  T2.0 $\pm$ 1.0 &   $>2.37       $   &    $3.09\pm0.10$ &   $0.52\pm0.05$	&   $0.37\pm0.07$  &   $0.89\pm0.07$	& 17.04   \\ 
SDSS J104829.21$+$091937.8  &  T2.5           &   $>2.81       $   &    $3.30\pm0.09$ &   $0.44\pm0.04$	&   $0.08\pm0.04$  &   $0.52\pm0.04$	& 16.39   \\ 
SDSS J143945.86$+$304220.6  &  T2.5           &   $>2.26       $   &    $3.27\pm0.10$ &   $0.44\pm0.04$	&   $0.19\pm0.06$  &   $0.63\pm0.06$	& 16.97   \\ 
 & \\
SDSS J120602.51$+$281328.7  &  T3.0           &   $>3.02       $   &    $3.38\pm0.08$ &   $0.27\pm0.04$	&  $-0.14\pm0.04$  &   $0.13\pm0.04$	& 16.10   \\ 
2MASS J12095613$-$1004008   &  T3.0           &   \nodata          &    \nodata       &   $0.31\pm0.04$	&   $0.07\pm0.04$  &   $0.38\pm0.04$	& 15.55      \\ 
SDSS J141530.05$+$572428.7  &  T3.0 $\pm$ 1.0 &   $>2.63       $   &    $3.38\pm0.09$ &   $0.40\pm0.04$	&   $0.54\pm0.04$  &   $0.94\pm0.04$	& 16.49   \\ 
SDSS J121440.95$+$631643.4  &  T3.5 $\pm$ 1.0 &   $>2.85       $   &    $3.60\pm0.10$ &   $0.25\pm0.04$	&   $0.07\pm0.05$  &   $0.32\pm0.05$	& 16.05   \\ 
SDSS J153417.05$+$161546.1AB\tablenotemark{d}  &  T3.5         &   $>2.30       $   &    $3.58\pm0.12$ &   $0.25\pm0.04$	&   $0.31\pm0.04$  &   $0.56\pm0.04$	& 16.62   \\ 
 & \\ 
SDSS J125011.65$+$392553.9  &  T4.0           &   $>2.55       $   &    $3.83\pm0.09$ &   $0.05\pm0.04$	&   $0.00\pm0.04$  &   $0.05\pm0.04$	& 16.12   \\ 
SDSS J135852.68$+$374711.9  &  T4.5 $\pm$ 1.0 &   $>2.61       $   &    $3.72\pm0.09$ &  $-0.24\pm0.04$	&  $-0.25\pm0.06$  &  $-0.49\pm0.06$	& 16.17   \\ 
SDSS J212413.89$+$010000.3  &  T5.0           &   $>2.79       $   &    $3.83\pm0.12$ &  $-0.24\pm0.04$	&  $ 0.05\pm0.04$  &  $-0.19\pm0.04$	& 15.88   \\ 
SDSS J032553.17$+$042540.1  &  T5.5           &   $>2.75       $   &    $3.85\pm0.09$ &  $-0.35\pm0.04$	&  $-0.22\pm0.05$  &  $-0.57\pm0.05$	& 15.88   \\ 
SDSS J163022.92$+$081822.0  &  T5.5           &   $>2.37       $   &    $3.95\pm0.10$ &  $-0.17\pm0.04$	&  $-0.06\pm0.04$  &  $-0.23\pm0.04$	& 16.18   \\ 
 & \\
2MASS J00345157$+$0523050\tablenotemark{e}   &  T6.5                &   \nodata       &   \nodata       &  $-0.44\pm0.04$  &  $-0.41\pm0.04$  &  $-0.85\pm0.04$	&    15.11    \\ 
SDSS J150411.63$+$102718.4  &  T7.0         &   $>2.16       $   &    $3.85\pm0.14$ &  $-0.43\pm0.04$	&  $-0.10\pm0.04$  &  $-0.53\pm0.04$	& 16.49   \\ 
SDSS J162838.77$+$230821.1  &  T7.0         &   $>2.27       $   &    $3.98\pm0.10$ &  $-0.38\pm0.05$	&  $-0.09\pm0.05$  &  $-0.47\pm0.04$	& 16.25   \\ 
\enddata
\tablenotetext{a}{ Uncertainties in spectral types are $\pm 0.5$ subtype, unless otherwise noted.}
\tablenotetext{b}{ $i-z$ colors of dwarfs not detected in $i$ are calculated using $5\sigma$ detection limit of $i = 22.5$, and indicated with ``$>$'' as lower limits.}
\tablenotetext{c}{ $J$ photometry errors are typically 0.03 mag, also refer to Table 1.}
\tablenotetext{d}{ Identified by \citealt{liu06} as a close binary with T1-T2 and T5-T6 components}.
\tablenotetext{e}{ Spectral type from Burgasser et al. (2005)}.
\end{deluxetable}
\clearpage

\end{document}